\documentclass[12pt,a4paper]{article}

\usepackage{amsthm}
\usepackage{amsmath,bbm}
\usepackage{graphicx,psfrag}
\usepackage[english]{babel}

\newcommand{\dx}{{\rm d}}
\newcommand{\Dt}[2]{\frac{\dx #1}{\dx #2}}
\newcommand{\Dtt}[2]{\frac{\dx^2  #1}{\dx #2^2}}
\newcommand{\Dttt}[2]{\frac{\dx^3 #1}{\dx #2^3}}
\newcommand{\e}{\varepsilon}

\newcommand{\Id}{\mathbbm{1}}
\newcommand{\E}{\mathbbm{E}}
\newcommand{\Pb}{\mathbbm{P}}
\newcommand{\Q}{\mathbbm{Q}}
\newcommand{\Or}{\mathcal{O}}

\newcommand{\R}{\mathbbm{R}}
\newcommand{\Z}{\mathbbm{Z}}
\newcommand{\N}{\mathbbm{N}}

\newenvironment{proofOF}[1]{\removelastskip\vspace{6pt}\noindent
 {\it Proof of #1.}}{\par\vspace{6pt} \qed}

\DeclareMathOperator*{\Ai}{Ai}

\numberwithin{equation}{section}

\newtheorem{theorem}{Theorem}[section]
\newtheorem{proposition}[theorem]{Proposition}
\newtheorem{lemma}[theorem]{Lemma}
\newtheorem{corollary}[theorem]{Corollary}
\newtheorem{definition}[theorem]{Definition}

\title{Scaling Limit for the Space-Time Covariance of the
Stationary Totally Asymmetric Simple Exclusion
Process}
\author{Patrik L.\ Ferrari and Herbert Spohn \\[6pt] {\normalsize Technische Universit\"at M\"unchen}\\ {\normalsize Zentrum Mathematik and Physik Department}\\ {\normalsize e-mails: ferrari@ma.tum.de, spohn@ma.tum.de}}
\date{28th September 2005}

\begin{document}
\maketitle
\sloppy

\begin{abstract}
The totally asymmetric simple exclusion process (TASEP) on the one-dimensional lattice with the Bernoulli $\rho$ measure as initial conditions, $0<\rho<1$, is stationary in space and time. Let $N_t(j)$ be the number of particles which have crossed the bond from $j$ to $j+1$ during the time span $[0,t]$. For $j=(1-2\rho)t+2w(\rho(1-\rho))^{1/3} t^{2/3}$ we prove that the fluctuations of $N_t(j)$ for large $t$ are of order $t^{1/3}$ and we determine the limiting distribution function $F_w(s)$, which is a generalization of the GUE Tracy-Widom distribution. The family $F_w(s)$ of distribution functions have been obtained before by Baik and Rains in the context of the PNG model with boundary sources, which requires the asymptotics of a Riemann-Hilbert problem. In our work we arrive at $F_w(s)$ through the asymptotics of a Fredholm determinant. $F_w(s)$ is simply related to the scaling function for the space-time covariance of the stationary TASEP, equivalently to the asymptotic transition probability of a single second class particle.
\end{abstract}

\section{Scaling limit and main result}\label{sect1} The totally asymmetric simple
exclusion process (TASEP) is, arguably, the simplest \emph{non-reversible}
interacting stochastic particle system. The occupation variables of the TASEP are
denoted by $\eta_j$, $j\in \Z$, $\eta_j=0$ means site $j$ is empty and $\eta_j=1$
means site $j$ is occupied. Since we plan to study the stationary space-time
covariance (= two-point function), the particles move on the entire
one-dimensional lattice $\Z$. The stochastic updating rule is extremely simple.
Particles jump to the right and are allowed to do so only if their right
neighboring site is empty. Jumps are independent of each other and are performed
after an exponential waiting time with mean 1, which starts from the time instant when
the right neighbor site is empty.

More precisely, we denote by $\eta$ a particle configuration, $\eta\in
\Omega=\{0,1\}^\Z$. Let $f$: $\Omega\to \R$ be a function depending only on a
finite number of $\eta_j$'s. Then the backward generator of the TASEP is given by
\begin{equation}\label{1.1}
Lf(\eta)=\sum_{j\in\Z}\eta_j(1-\eta_{j+1})\big(f(\eta^{j,j+1})-f(\eta)\big).
\end{equation}
Here $\eta^{j,j+1}$ denotes the configuration $\eta$ with the
occupations at sites $j$ and \mbox{$j+1$} interchanged. The semigroup $e^{Lt}$ is
well-defined as acting on bounded and continuous functions on $\Omega$. $e^{Lt}$
is the transition probability of the TASEP~\cite{Li99}.

Let $\mu_\rho$ be the Bernoulli measure with density $\rho$, $0\leq\rho\leq 1$,
i.e., under $\mu_\rho$ the $\eta_j$'s are independent and
$\mu_\rho(\eta_j=1)=\rho$. From (\ref{1.1}) it is easy to check that
\begin{equation}\label{1.2}
\mu_\rho(Lf)=0
\end{equation}
for all local functions $f$, which means that the Bernoulli measures are
stationary measures for the TASEP. In fact, these are the only translation
invariant stationary measures~\cite{Lig76}. In the sequel we fix $\rho$, excluding
the degenerate cases $\rho=0$, $\rho=1$, and start the TASEP with $\mu_\rho$. The
corresponding space-time stationary process is denoted by $\eta_j(t)$, $t\in\R$,
$j\in\Z$. $\Pb_{\rm TA}$ denotes the probability measure on paths $t\mapsto\eta(t)$ and $\E_{\rm TA}$ its expectation. The dependence on $\rho$ is always understood implicitly. Note that the average current for the stationary TASEP is $j(\rho)=\rho(1-\rho)$.

As for any other stationary stochastic field theory the most basic quantity is the
two-point function, which for the TASEP is defined through
\begin{equation}\label{1.3}
\E_{\rm TA}\big(\eta_j(t)\eta_0(0)\big)-\rho^2=S(j,t).
\end{equation}
For fixed $t$, $S(j,t)$ decays exponentially in $j$. One has the sum rules
\begin{equation}\label{1.4}
\sum_{j\in\Z}S(j,t)=
\sum_{j\in\Z}\big(\E_{\rm TA}(\eta_j(t)\eta_0(0))-\rho^2\big)
=\rho(1-\rho)=\chi(\rho),
\end{equation}
\begin{equation}\label{1.5}
\frac{1}{\chi}\sum_{j\in\Z}j
S(j,t)=j'(\rho)t=(1-2\rho)t.
\end{equation}
$\chi^{-1}S(j,t)$ can be viewed as the probability for a second class particle to be at site $j$ at time $t$ given it was at $j=0$ initially, see e.g.~\cite{PS01}. Thus
\begin{equation}\label{1.6}
S(j,t)\geq 0,
\end{equation}
which would not hold on general grounds.

The next finer information is the variance
\begin{equation}\label{1.7}
\sigma(t)^2=\chi^{-1}\sum_{j\in\Z}j^2
S(j,t)-((1-2\rho)t)^2.
\end{equation}
Naively, one might expect that $\sigma(t)\cong\sqrt{t}$, arguing that the second class particle moves random walk like. As noticed in~\cite{Sp83}, in a purely perturbative argument, $\sigma(t)$ is likely to grow faster than $\sqrt{t}$. The proper scaling form was firmly established in~\cite{BKS85} with the result
\begin{equation}\label{1.8}
\sigma(t)\cong a_0 \chi^{1/3}t^{2/3}
\end{equation}
for large $t$. $\chi^{1/3}$ follows on dimensional grounds, while the prefactor $a_0$ has to be determined numerically. In fact, $a_0= 2.0209\ldots$ which is a consequence of the result reported here together with~\cite{PS02b}.

Forster, Nelson and Stephen~\cite{FNS77} consider, as a particular case of the fluctuating Navier-Stokes equation, the stochastic Burgers equation
\begin{equation}\label{1.9}
\frac{\partial}{\partial t}u= \frac{\partial}{\partial x}
\big(-u^2+\nu \frac{\partial}{\partial x}u+\xi\big)
\end{equation}
with $\nu>0$ and $\xi$ space-time white noise, which is a sort of
continuum stochastic partial differential equation version of the TASEP. They obtain the dynamical exponent $z=3/2$ which corresponds to the $2/3$ of (\ref{1.8}). Kardar, Parisi and Zhang \cite{KPZ86} study surface growth which for a
one-dimensional substrate reduces to (\ref{1.9}) with $u$ being the gradient of the height function. By more refined arguments they confirm $z=3/2$ in one space dimension. Since then many approximate theories have appeared, see e.g.~\cite{PS01,KMH92} for a more complete discussion. The only one which survives the test is the mode-coupling theory, which is a nonlinear equation for $S(j,t)$ \cite{BKS85}. A careful, rather recent, numerical study~\cite{CM02} of this equation yields surprisingly good agreement with the exact two-point function in the scaling limit~\cite{PS02b}.

The power law (\ref{1.8}) strongly suggests the scaling form
\begin{equation}\label{1.10}
S(j,t)\cong \frac{\chi}{4}(2\chi^{1/3}t^{2/3})^{-1} g''_{\textrm{sc}}\big((j-(1-2\rho)t) (2\chi^{1/3}t^{2/3})^{-1}\big)
\end{equation}
for large $t$ and for $j-(1-2\rho)t=\mathcal{O}(t^{2/3})$ with the
scaling function $g''_{\textrm{sc}}/4$ independent of $\rho$. Our
main result will be to prove a version of (\ref{1.10}) with a
reasonably explicit expression for $g''_{\textrm{sc}}$.

The scaling function $g_{\textrm{sc}}$ appeared already in the context of the
variance of the height differences in the polynuclear growth (PNG) model, where
$g_{\textrm{sc}}$ is somewhat indirectly determined by a set of differential
equations, which were discovered by Baik and Rains~\cite{BR00}, see Appendix~\ref{AppBRlink}. These differential equations are solved numerically in~\cite{PS02b}, where also a plot of $g''_{\textrm{sc}}$ is displayed. Thus one important consequence of our main result is to establish that in the scaling limit the PNG model and the TASEP have the same scaling function for their covariance. Such a property is expected for a much larger class of one-dimensional driven lattice gases. For example, if instead of the TASEP we allow for partial asymmetry, to say a particle jumps with probability $p$ to the right and $1-p$ to the left, $p\neq 1/2$, then (\ref{1.10}) should still hold provided $(1-2\rho)t$ is replaced by $(2p-1)(1-2\rho)t$. The general formulation of the universality hypothesis for one-dimensional driven lattice gases is explained in~\cite{KMH92}, see also~\cite{PS01}. Viewed in this context our main result asserts that the TASEP and the PNG model are in the same universality class.

The issue of universality is certainly one strong motivation for our study. At
first sight PNG and TASEP look very different, while when viewed properly they are in fact not so far apart. The interpolating family of models is the TASEP with a discrete time updating rule. Its extreme limits are the PNG model on one side and the continuous time TASEP on the other side, see~\cite{PS01}. Given the scaling limit for the PNG, it is not surprising to have the same result for the TASEP. However, it turns out that the method used in proving the analogue of (\ref{1.10}) for the PNG model does not generalize to the TASEP, which for us is an even more compelling reason to investigate the TASEP. At a certain stage the proof in~\cite{PS02b} uses that the two-dimensional Poisson process is invariant under linear scale changes, which is a property special to the PNG model. For the TASEP we have to develop a novel method which will be rather different from~\cite{BR00,PS02b} and uses non-intersecting line ensembles. In fact, our expression for $g''_{\textrm{sc}}$ has an appearance quite unlike to the one discovered by Baik and Rains. It requires an argument that both expressions are in agreement, see Appendix~\ref{AppBRlink}.

To state our main result we have to first reformulate the TASEP as
a growth process by introducing the height function $h_t(j)$ through
\begin{equation}\label{1.11}
h_t(j)=
 \begin{cases}
 2N_t +\sum^j_{i=1}(1-2\eta_i(t)) & \textrm{for }j\geq 1,\\
 2N_t & \textrm{for }j=0,\\
 2N_t -\sum^0_{i=j+1}(1-2\eta_i(t)) & \textrm{for }j\leq -1,
 \end{cases}
\end{equation}
$t\geq 0$, where $N_t$ counts the number of jumps from site 0 to
site 1 during the time-span $[0,t]$. Note that $h_t(j)-h_0(j)=2 N_t(j)$, where $N_t(j)$ counts the number of particles which have crossed the bond from $j$ to $j+1$ during the time span $[0,t]$. By stationarity one has
\begin{equation}\label{1.12}
\E_{\rm TA}(h_t(j))=2\rho(1-\rho) t+(1-2\rho)j.
\end{equation}

Since $h_t(j+1)-h_t(j)=-2\eta_{j+1}(t)$, the variance of the
height must be simply related to $S(j,t)$.
\begin{proposition}\label{Prop1}
Let $\Delta$ be the discrete Laplacian, $(\Delta f)(j)=f(j+1)+f(j-1)-2f(j)$. Then
\begin{equation}\label{1.13}
8 S(j,t)=\big(\Delta\E_{\rm TA}\big([h_t(\cdot) - \E_{\rm TA}(h_t(\cdot))]^2\big)\big)(j).
\end{equation}
\end{proposition}
\noindent The proof can be found in Proposition 4.1 of~\cite{PS01}.

We introduce the family of distribution functions\footnote{Our $F_w$ equals the $F_{w/2}$ in the definition in Conjecture 7.2 of~\cite{PS01}. The reason for this change is that it slightly simplifies the functions below and also they match with the choice in~\cite{SI04}.}
\begin{eqnarray}\label{1.14}
 F_w(s,t)& = & \Pb_{\rm TA}\big(\{(1-2\chi)t+2w(1-2\rho)\chi^{1/3}t^{2/3}
-2s\chi^{2/3}t^{1/3}\nonumber\\
&  &\phantom{\Pb_{\rm TA}\big(} \leq  h_t(\lfloor(1-2\rho)t+2w\chi^{1/3}t^{2/3}\rfloor)\}\big),
\end{eqnarray}
where $\lfloor x\rfloor$ denotes the integer part of $x$. Here the height is evaluated at $(1-2\rho)t$, which is determined by the propagation of a tiny density fluctuation, plus a in comparison small off-set of order $t^{2/3}$, while the distribution function is centered at $\E_{\rm TA}(h_t(j))$ with $j=\lfloor(1-2\rho)t+2w\chi^{1/3}t^{2/3}\rfloor$ and has an argument, $-s$, which lives on the scale $\chi^{2/3}t^{1/3}$.

As to be shown, the distribution function $F_w(s,t)$ converges to a limit as $t \to \infty$. The limit will be expressed in terms of a scaling function
$g$ and the GUE Tracy-Widom distribution function $F_{\rm GUE}(s)$. The latter can be written as a Fredholm determinant in $L^2(\R)$,
\begin{equation}
F_{\rm GUE}(s)=\det(\Id-P_0 K_{\Ai,s} P_0)
\end{equation}
with $P_0$ the projector operator on $[0,\infty)$ and $K_{\Ai,s}$ the integral operator with the Airy kernel shifted by $s$, i.e., \begin{equation}
K_{\Ai,s}(x,y)=\int_{\R_+}\dx\lambda \Ai(\lambda+x+s)\Ai(\lambda+y+s).
\end{equation}
Define the functions
\begin{eqnarray}
 \widehat \Phi_{w,s}(x) &=& \int_{\R_-}\dx z e^{w z} K_{\Ai,s}(z,x) e^{w s}, \nonumber \\
 \widehat \Psi_{w,s}(y) &=& \int_{\R_-}\dx z e^{w z} \Ai(y+z+s), \\
 \rho_s(x,y) &=& (\Id-P_0 K_{\Ai,s} P_0)^{-1}(x,y), \nonumber
\end{eqnarray}
and the scaling function $g$ by
\begin{eqnarray}\label{eqScalingG}
g(s,w)&=&e^{-\frac13 w^3} \bigg[\int_{\R_-^2}\dx x \dx y e^{w(x+y)}\Ai(x+y+s)\nonumber \\
& &\hspace{26pt}+\int_{\R_+^2}\dx x\dx y \widehat \Phi_{w,s}(x)\rho_s(x,y)\widehat \Psi_{w,s}(y)\bigg].
\end{eqnarray}
Our main theorem asserts the limit of the family of distribution functions $F_w(s,t)$.
\begin{theorem}\label{MainThm}
Let $F_{\rm GUE}$ and $g$ defined above. Then for fixed $c_1 < c_2$ one has
\begin{equation}\label{1.15}
\lim_{t\to\infty} \int_{c_1}^{c_2} F_w(s,t) \dx s= F_{\rm GUE}(c_2+w^2)g(c_2+w^2,w) - F_{\rm GUE}(c_1+w^2)g(c_1+w^2,w)
\end{equation}
pointwise.
\end{theorem}

\begin{corollary}\label{corollario}
The limiting height distribution function $F_w(s)$ is given by
\begin{equation}\label{eq6.11}
F_w(s)=\frac{\partial}{\partial s}\big(F_{\rm GUE}(s+w^2) g(s+w^2,w)\big).
\end{equation}
\end{corollary}
For the PNG model Baik and Rains obtain the limiting height distribution function denoted by $H(s+w^2;w/2,-w/2)$ in Defintion 3 of~\cite{BR00}. It has the same structure as $F_w(s)$. Only the scaling function $g$ is given as the solution of a set of differential equations, see Appendix~\ref{AppBRlink}.

The two-point function of the TASEP carries information on the variance of height differences, see (\ref{1.13}), while Theorem~\ref{MainThm} provides the full family of distribution functions. In this sense (\ref{1.15}) is a stronger result than (\ref{1.10}). On the other hand, the limit (\ref{1.15}) for the distribution function asserts only the weak convergence of the corresponding probability measures, while from (\ref{1.13}) we infer that for the space-time covariance the convergence of second moments would be needed. If we \textit{assume} a suitable tightness condition on $F_w(s,t)$, then
\begin{equation}\label{1.16}
\lim_{t\to\infty} \int s^2 \dx F_w(s,t)= \int s^2 \dx
F_w(s)=g_\textrm{sc}(w),
\end{equation}
which together with Proposition \ref{Prop1} yields
\begin{equation}\label{1.17}
\lim_{t\to\infty} 2\chi^{1/3} t^{2/3} S(\lfloor(1-2\rho)t
+2w\chi^{1/3}t^{2/3}\rfloor,t) =\frac{1}{4}\chi g''_{\rm sc}(w)
\end{equation}
when integrated against an arbitrary smooth function in $w$, in agreement
with the claim (\ref{1.10}). Tightness is also missing in the analysis of the PNG model.

Over the recent years there has been a considerable interest in scaling limits for the TASEP. Slightly more general than here, one considers an initial measure which is Bernoulli $\rho_-$ in the left half lattice $\Z_-$ and Bernoulli $\rho_+$ in the right half lattice $\Z_+$. The initial step, $\rho_-=1$ and $\rho_+=0$, is studied by Johansson~\cite{Jo00b} by mapping the TASEP to a last passage percolation problem. For general $\rho_+$ and $\rho_-$ such a map is still possible and yields a last passage percolation problem with boundary conditions~\cite{PS01}. Through the Robinson-Schensted-Knuth (RSK) correspondence one then obtains a line ensemble with boundary sources. This line ensemble is determinantal, in fact only a rank one perturbation of the line ensemble with tie-down at both ends. We refer to~\cite{SI04}, where a similar construction is carried out for the line ensemble corresponding to the discrete time TASEP. There is also a link to the work by Baik, Ben Arous, and P\'ech\'e~\cite{BBP04}, who study rank $r$ perturbations of the complex Gaussian sample covariance matrices. Viewed from this perspective the stationary TASEP is singular, which is partially overcome by the shift argument, see also~\cite{SI04}. But even
then, in the resulting matrix elements there is still a delicate cancellation which tends to hide the asymptotics. The technique of line ensembles can be used also for the investigation of multi-point statistics~\cite{PS02,SI04}, which however will not be needed in our context.

In computer simulations mostly deterministic flat initial conditions are adopted, which translates to the initial particle configuration $\ldots 010101\ldots$ of the TASEP. As established by Sasamoto~\cite{Sas05}, see also~\cite{FS05b}, the single point statistics in the limit of large times is then given by the distribution of the largest eigenvalue of the GOE of random matrices and thus different from the distribution obtained in this contribution. For the PNG model the corresponding result is proved prior by Baik and Rains~\cite{BR99}, see also~\cite{Fer04}.

Our paper is divided into two parts. The first part is a fixed $t$ discussion of $F_w(s,t)$ with the goal to obtain a manageable expression. The second part is devoted to the asymptotic analysis. In the Appendices we establish that our expression for $F_w(s)$ agrees with the one of Baik and Rains, provide some background on the determinantal fields turning up, and explain how the line ensemble is related to the Laguerre kernel.

\section*{Acknowledgements.} This project started in the fall 2003 at the Newton Institute workshop ``Interaction and Growth in Complex Stochastic Systems". Michael Pr\"{a}hofer explained to HS the shift argument, which opened the route for using Fredholm determinants. A further important input was an early 2003 note from J.~Baik, which indicated that boundary sources generate a rank one perturbation. HS is grateful for both contributions. We also greatly benefited from illuminating discussions with T.~Sasamoto. E.~Carlen helped us with the proof of Proposition~\ref{PropInverse}. 

\section{Map to a directed polymer}\label{sect2}
The statistics of the height function $h_t(j)$, restricted to the cone \mbox{$\{j,h||j|\leq h\}$}, can be represented through a directed last passage percolation, see~\cite{PS01}. For this purpose, in the initial configuration, let $\zeta_+ +1$ be the location of the first particle to the right of (and including) 1 and let $-\zeta_-$ be the location of the first hole to the left of (and including) 0. Therefore $\zeta_-,\zeta_+$ are independent and geometrically distributed, $\Q(\zeta_-=n)=(1-\rho)\rho^n$, $\Q(\zeta_+=n)=\rho(1-\rho)^n$, $n=0,1,\ldots$. In addition we define the family of independent exponentially distributed random variables $w(i,j)$, $i,j\geq 0$, such that $w(i,j)$ has mean $1$ for $i,j\geq 1$, $w(i,0)$ has mean $(1-\rho)^{-1}$ for $i\geq 1$, $w(0,j)$ has mean $\rho^{-1}$ for $j\geq 1$, and $w(0,0)=0$. The joint distribution of the random variables $\zeta = (\zeta_+,\zeta_-)$ and $\{w(i,j), i,j \geq 0\}$ is
denoted by $\Q$. These exponentially distributed random variables are linked to the TASEP in the following way: $w(\zeta_+ +\ell,0)$ is the $\ell$-th waiting time of the first particle to the right of 0 and $w(0,\zeta_- +\ell)$ is the $\ell$-th waiting time of the first hole to the left of 0, $\ell=1,2,\ldots$. To describe the other $w(i,j)$'s we label in the initial configuration the particles from right to left such that the first particle to the right of 0 has label 0. Then $w(i,j)$, $i,j\geq 1$, is the $j$-th waiting time of particle $i$, where the first waiting time refers to the instant when the $i$-th particle is at lattice site $-i+1$.

For given $\zeta$ let
\begin{equation}\label{3.1}
w_\zeta(i,j)=
 \begin{cases}
 0 & \textrm{for }1\leq i\leq \zeta_+,j=0,\\
 0 & \textrm{for }i=0,1\leq j\leq \zeta_-,\\
 w(i,j) & \textrm{otherwise}.
 \end{cases}
\end{equation}
The $w_\zeta(i,j)$'s are used as local passage times in a directed last passage percolation. Let us consider an up/right path $\omega$ on $\N^2$ with a finite number of steps. To it we assign the passage time
\begin{equation}\label{3.2}
T(\omega)=\sum_{(i,j)\in\omega}w_\zeta(i,j).
\end{equation}
Then the last passage time from point $(0,0)$ to point $(m,n)$ is given by
\begin{equation}\label{3.3}
G(m,n)= \max_{\omega:(0,0)\to(m,n)}T(\omega).
\end{equation}
Here the maximum is over the up/right paths which start at $(0,0)$ and end at $(m,n)$.

\begin{proposition}\label{Prop2} {\rm \cite{PS01}}
With the above definitions
\begin{equation}\label{3.4}
\Q(\{G(m,n)\leq t\})=\Pb_{\rm TA}(\{m+n\leq h_t(m-n)\}).
\end{equation}
\end{proposition}

$G(m,n)$ can also be viewed as a growth process. We introduce the corresponding height function $\tilde{h}(j,\tau)$, $j\in\Z$, $\tau\in\N$, through
\begin{eqnarray}\label{3.5}
\tilde{h}(j,\tau)&=&0,\textrm{ for } |j|\geq \tau,\\
\tilde{h}(j,\tau)&=&
 \begin{cases}
 G((\tau-1+j)/2,(\tau-1-j)/2), & \textrm{if }(-1)^{\tau+j}=-1,\\
 G((\tau-2+j)/2,(\tau-2-j)/2), & \textrm{if }(-1)^{\tau+j}=1,
 \end{cases}\nonumber \\
& &\phantom{0,}\textrm{ for }|j|<\tau. \nonumber
\end{eqnarray}
By Proposition \ref{Prop2},
\begin{equation}\label{3.6}
\Q\big(\{\tilde{h}(j,\tau)\leq t\}\big)=
 \begin{cases}
 \Pb_{\rm TA}\big(\{\tau-1\leq h_t(j)\}\big), & \textrm{if } (-1)^{\tau+j}=-1,\\
 \Pb_{\rm TA}\big(\{\tau-2\leq h_t(j)\}\big), & \textrm{if }(-1)^{\tau+j}=1.
 \end{cases}
\end{equation}

$\tilde{h}(j,\tau)$ is not such a convenient quantity and we modify it by allowing an error of order 1. We display the $\zeta$-dependence of $G(m,n)$ explicitly through $G^{\zeta}(m,n)$. In particular, $G^0(m,n)$ is the random variable obtained by setting $\zeta_+=0=\zeta_-$.

\begin{proposition}\label{Prop3}
Uniformly in the endpoint one has
\begin{equation}\label{3.7}
\Q\big(\{|G^\zeta(m,n)-G^0(m,n)|\geq u\} | \zeta\big)\leq \zeta_+ e^{-u/(1-\rho)}+\zeta_- e^{-u/\rho}.
\end{equation}
\end{proposition}
\begin{proof} We fix the endpoint $(m,n)$. Let $T^\zeta(\omega)$ be the passage time for $w_\zeta(i,j)$ and let $\omega^\zeta_{\max}$ be a maximizing path from $(0,0)$ to $(m,n)$. Then $G^\zeta(m,n)=T^\zeta(\omega^\zeta_{\max})$, $G^0(m,n)=T^0(\omega^0_{\max})$. One has
\begin{eqnarray}\label{3.8}
G^0(m,n)-G^\zeta(m,n)&=&
T^0(\omega^0_{\max})-T^\zeta(\omega^0_{\max})
+T^\zeta(\omega^0_{\max})-T^\zeta(\omega^\zeta_{\max})\nonumber\\
&\leq &T^0(\omega^0_{\max})-T^\zeta(\omega^0_{\max})\nonumber\\
&\leq &\sum^{\zeta_+}_{i=1} w(i,0) +\sum^{\zeta_-}_{j=1} w(0,j),
\end{eqnarray}
where in the first inequality we used that $\omega^\zeta_{\max}$ is a
maximizer of $T^\zeta$. Similarly
\begin{eqnarray}\label{3.9}
G^0(m,n)-G^\zeta(m,n)&=&
T^0(\omega^0_{\max})-T^0(\omega^\zeta_{\max})
+T^0(\omega^\zeta_{\max})-T^\zeta(\omega^\zeta_{\max})\nonumber\\
&\geq &T^0(\omega^\zeta_{\max})- T^\zeta(\omega^\zeta_{\max})\geq
0.
\end{eqnarray}
Combining (\ref{3.8}), (\ref{3.9}) yields
\begin{eqnarray}\label{3.10}
\Q\big(\{|G^\zeta(m,n)-G^0(m,n)|\geq u\}|\zeta\big)&\leq & \Q\Big(\{\sum^{\zeta_+}_{i=1} w(i,0) +
\sum^{\zeta_-}_{j=1} w(0,j)\geq u\}|\zeta \Big)\nonumber\\
&\leq &\zeta_+ e^{-u/(1-\rho)} +\zeta_- e^{-u/\rho}.
\end{eqnarray}
\end{proof}

\begin{definition}\label{Definh0}
$h^0(j,\tau)$ is the height function as given in (\ref{3.5}), where $G(m,n)$ is replaced by $G^0(m,n)$ with the corresponding passage times $w_0(i,j)$.
\end{definition}

It follows from the identity (\ref{3.6}) and Proposition
\ref{Prop3} that for $\e>0$, independent of $t$, one can choose
$d(\e)$ such that
\begin{equation}\label{3.11}
(1-\e)\Q\big(\{h^0(j,\tau)\leq t-d(\e)\}\big)\leq F_w(s,t)\leq
\Q\big(\{h^0(j,\tau)\leq t+d(\e)\}\big)+\e,
\end{equation}
uniformly in $j,\tau$. Therefore, setting
\begin{eqnarray}\label{3.12}
F^0_w(s,t)&=&\Q\big(\{h^0(\lfloor (1-2\rho)t+2w\chi^{1/3}t^{2/3}\rfloor, \\
& & \lfloor(1-2\chi)t+2w(1-2\rho)\chi^{1/3}t^{2/3}-2 s\chi^{2/3}t^{1/3}\rfloor)\leq t\}\big), \nonumber
\end{eqnarray}
one has
\begin{equation}\label{3.13}
(1-\e)F^0_w\big(s,t-d(\e)\big)\leq F_w(s,t)\leq
F^0_w\big(s,t+d(\e)\big)+\e.
\end{equation}
Since $t\to\infty$, we conclude that Theorem~\ref{MainThm} is implied by Theorem~\ref{Thm2} stated below.

\begin{theorem}\label{Thm2}
For fixed $c_1<c_2$ the following limit holds
\begin{equation}\label{3.14}
 \lim_{t\to\infty} \int_{c_1}^{c_2} F^0_w(s,t)\dx s =\int_{c_1}^{c_2} F_w(s)\dx s.
\end{equation}
\end{theorem}
\noindent The remainder of the paper deals with the proof of Theorem~\ref{Thm2}.

\section{The Laguerre line ensemble with boundary sources}\label{sect3}
Let us consider the directed polymer in the general case of independent $w(i,j)$'s with exponential distribution of mean $1/a_{ij}$, $i,j\geq 0$. The directed polymer is determinantal (a notion which will be explained below) provided $a_{ij}=a_i+b_j>0$. For the case of Theorem~\ref{Thm2} one has to deal with $w_0(i,j)$ and the obvious choice would be $a_i=\frac{1}{2}-(1-\rho)\delta_{i,0}$, $b_j=\frac{1}{2}-\rho\delta_{j,0}$. The corresponding directed polymer fails to be determinantal on two accounts:\smallskip\\
(i) $a_0+b_0=0$ whereas it should be striclty positive,\smallskip\\
(ii) formally $w(0,0)$ is uniformly distributed on $\R_+$, while in actual fact $w_0(0,0)=0$.

Our strategy is to first discuss the line ensemble for
$a_i=\frac{1}{2}+(a-\frac{1}{2})\delta_{i,0}$,
$b_j=\frac{1}{2}+(b-\frac{1}{2})\delta_{j,0}$, $a>0$, $b>0$. This
is the task of the current section. In the following section we
will show that the case $w(0,0)=0$ can be deduced from a
shift argument. The resulting expressions will then be analytically
continued to $a=\rho-\frac{1}{2}$, $b=\frac{1}{2}-\rho$. In this limit we recover $w_0(i,j)$, which is required for Theorem~\ref{Thm2}.

\begin{definition}\label{defin3.1}
Let $w_{a,b}(i,j)$, $i,j\in\N$, be a family of independent exponentially distributed random variables such that
\begin{equation}
\E(w_{a,b}(i,j))^{-1}=1+(a-\tfrac12)\delta_{i,0}+(b-\tfrac12)\delta_{j,0}
\end{equation}
with $0<a,b<\tfrac12$.
\end{definition}

With $w_{a,b}(i,j)$ as in Definition~\ref{defin3.1} let $T(\omega)$ be as in (\ref{3.2}) with $w_\zeta(i,j)$ replaced by $w_{a,b}(i,j)$ and let
\begin{equation}\label{4.1}
G(m,n)=\max_{\omega:(0,0)\to(m,n)} T(\omega),
\end{equation}
compare with (\ref{3.3}). We define the height function
$h(j,\tau)$, $j\in\Z$, $\tau\in\N$, through
(\ref{3.5}). It can also be generated by the following growth
process,
\begin{eqnarray}\label{4.3}
h(j,0)&=&0, \\
h(j,\tau+1)&=& \left\{
\begin{array}{ll}
 \max\{h(j-1,\tau),h(j+1,\tau)\}& \\
 \phantom{\max}+w_{a,b}((\tau+j)/2,(\tau-j)/2), & \textrm{if }(-1)^{j+\tau}=1,\\[6pt]
 h(j,\tau), & \textrm{if }(-1)^{j+\tau}=-1,
\end{array}\right. \nonumber \\
& & \phantom{0,\,}\textrm{ for }|j|<\tau+1,\nonumber \\
h(j,\tau+1)&=&0, \textrm{ for }|j|\geq\tau+1.\nonumber
\end{eqnarray}

\begin{figure}[t!]
\begin{center}
\psfrag{1}[c]{$\tau=1$}
\psfrag{2}[c]{$\tau=2$}
\psfrag{3}[c]{$\tau=3$}
\includegraphics[width=\textwidth]{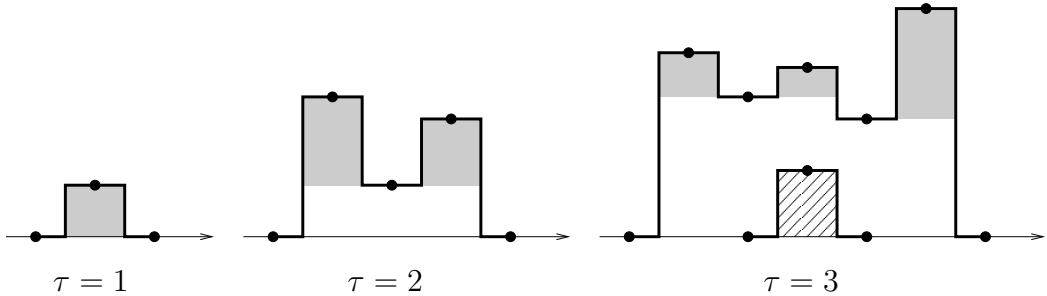}
\caption{The growth dynamics associated with the TASEP directed
last passage percolation.}\label{Fig1}
\end{center}
\end{figure}

The dynamics is best visualized by extending $h(j,\tau)$ to a function over $\R$
through $h(x,\tau)=h(j,\tau)$ for $j-\frac{1}{2}\leq x<j+\frac{1}{2}$, see
Figure~\ref{Fig1}.
Then alternately there is a stochastic and deterministic up-date. In the
stochastic up-date mass is added to the current height $h(x,\tau)$ according to
$w_{a,b}(i,j)$, see (\ref{4.3}). In the deterministic up-date down-steps move one unit
to the right and up-steps one unit to the left. Thereby parts of the up-dated $h$
may overlap. The maximum rule means that the excess mass in the overlap is
annihilated.

Underlying the growth process one may construct the corresponding Robinson-Schensted-Knuth (RSK) dynamics~\cite{Jo03b}, which in our case simply means that the overlap annihilated in line $\ell$ is copied to the lower lying line $\ell-1$. In formulas we set
\begin{eqnarray}\label{4.4}
h_0(j,\tau)&=&h(j,\tau),\nonumber \\
h_\ell(j,0)&=&0,\\
h_{\ell-1}(j,\tau+1)&=&
\left\{\begin{array}{ll}
h_{\ell-1}(j,\tau)-h_\ell(j,\tau)& \\
+\min\{h_\ell(j-1,\tau), h_\ell(j+1,\tau)\}, & \textrm{if }(-1)^{\tau+j}=1,\\[6pt]
 h_{\ell-1}(j,\tau), & \textrm{if }(-1)^{\tau+j}=-1,
\end{array}\right. \nonumber
\end{eqnarray}
with the line label $\ell=0,-1,\ldots$.

The purpose of the RSK construction consists in having, for fixed $\tau$, a
manageable statistics of the collection of points $\{h_\ell(j,\tau)$,
$\ell\in\Z_-$, $|j|<\tau$, $h_\ell(j,\tau)>0\}$. To describe their statistics
directly without recourse to the stochastic dynamics we first have to define admissible point configurations. Let $\{x_j, j=-n,\ldots,0\}$ be points on $[0,\infty)$ ordered as $0\leq x_{-n}\leq\ldots\leq x_0$. We say that $\{x_j, j=-n,\ldots,0\}\prec\{x_j', j=-n,\ldots,0\}$ if
$x_0\leq x'_0$, $x_j \leq x'_j\leq x_{j+1}$ for $j=-n,\ldots,-1$. Admissible point
configurations are then
\begin{eqnarray}\label{4.5}
h_\ell(\pm\tau,\tau)&=&0, \\
\{h_\ell(j,\tau),\ell\in\Z_-\}&\prec&
\{h_\ell(j+1,\tau),\ell\in\Z_-\} \textrm{ if }|j|<\tau
\textrm{ and }(-1)^{j+\tau}=-1,\nonumber \\
\{h_\ell(j,\tau),\ell\in\Z_-\} &\succ&
\{h_\ell(j+1,\tau),\ell\in\Z_-\} \textrm{ if }|j|<\tau
\textrm{ and }(-1)^{j+\tau}=1. \nonumber
\end{eqnarray}

\begin{figure}[t!]
\begin{center}
\psfrag{0}[c]{$0$}
\includegraphics[height=4cm]{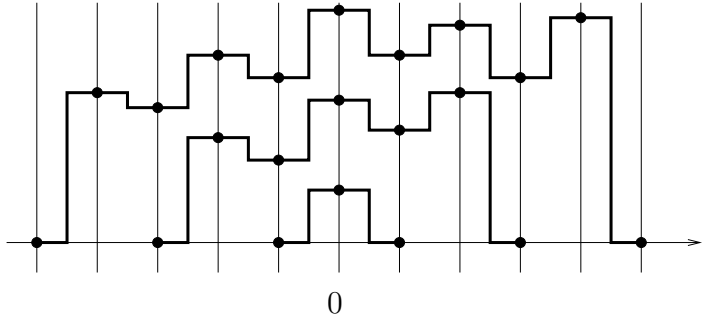}
\caption{A non-intersecting line ensemble at $\tau = 5$.}\label{Fig2}
\end{center}
\end{figure}

As with the growth dynamics, the order $\prec$ and $\succ$ can be
visualized by extending $h_\ell(j,\tau)$ to $\R$ by
setting $h_\ell(x,\tau)=h_\ell(j,\tau)$ for $j-\frac{1}{2}\leq
x<j+\frac{1}{2}$. Then (\ref{4.5}) means that the lines
$h_\ell(x,\tau)$ do not intersect when considered as lines in the
plane, see Figure~\ref{Fig2}.

To a given point configuration, \emph{alias} line ensemble, one associates a weight. It is the product of the weights for each single jump. Let us use $\delta$ as the generic symbol for a height difference. Then the up-step $h_0(-\tau,\tau)$ to $h_0(-\tau+1,\tau)$ has weight $e^{-b|\delta|}$ and the down-step $h_0(\tau-1,\tau)$ to $h_0(\tau,\tau)$ has weight $e^{-a|\delta|}$. All other jumps of the form
$h_\ell(j,\tau)$ to $h_\ell(j+1,\tau)$ have weight $e^{-|\delta|/2}$. Note that the weights are assigned by reading the vector $\vec{a}$ from right to left and the vector $\vec{b}$ from left to right. The total weight is normalized to become a probability. This probability measure is called the \textit{Laguerre line ensemble} with boundary values $a,b$. It agrees with the probability measure at growth time $\tau$ obtained from the growth dynamics (\ref{4.4}) together with the RSK construction.

It is convenient to think of $\{h_\ell(j,\tau),\ell\in\Z_-,j\in\Z\}$ as a point process on $\Z\times(0,\infty)$, where $j$ is referred to as time and $h_\ell$ as space. The corresponding random field is then
\begin{equation}\label{4.6}
\phi_\tau(j,y)=\sum_{\ell \leq 0}\delta(h_\ell(j,\tau)-y),\quad y>0.
\end{equation}
According to our construction, at $\Z\times\{0\}$, i.e.\ at $y=0$, there are an infinite number of points. However, the point measure refers only to points with a strictly positive $y$ coordinate. In fact, $\phi_\tau(j,y)$ is supported by
\begin{equation}\label{4.7}
\sum^{\lfloor(2\tau-1)/4\rfloor}_{j=0} (2\tau-4j-1)
\end{equation}
points, $\lfloor\cdot\rfloor$ denoting the integer part. The point process $\phi_\tau(j,y)$ is \emph{determinantal}, in the sense that is has determinantal moments. This
means that there exists a kernel $K^{(\tau)}_{a,b}(j,y;j',y')$ such
that for a time-ordered sequence $j_1\leq \ldots\leq j_m$ and
arbitrary space-points $y_1,\ldots,y_m>0$ the $m$-th moment of
$\phi_\tau$ is given by
\begin{equation}\label{4.8}
\E\Big(\prod^m_{k=1}\phi_\tau(j_k,y_k)\Big)=
\det\big(K^{(\tau)}_{a,b}(j_k,y_k;j_{k'},y_{k'})\big)_{1\leq k,k'\leq m}.
\end{equation}

For the two-point function of the TASEP we need the statistics of the random field $\phi_\tau(j,y)$ only at fixed time $j$ and in the remainder of this section we will provide an expression for $K^{(\tau)}_{a,b}(j,y;j,y')$, $y,y'>0$. There is no difficulty in principle to extend the construction and our results to unequal times.

The distinction between odd and even $j+\tau$ is slightly cumbersome and we restrict to odd $\tau$, even $j$ by setting
\begin{equation}\label{4.9}
\tau=2m+1, \quad j=2d.
\end{equation}

In $L^2(\R)$ we define $P_+$ as projection onto $\R_+$, $P_+ +P_-=1$. We also introduce the operators $T_+$, $T_-$ with integral kernels
\begin{eqnarray}\label{4.10}
T_+(x,y)&=&e^{-(x-y)/2}\Theta(x-y),\nonumber\\
T_-(x,y)&=&e^{-(y-x)/2}\Theta(y-x),
\end{eqnarray}
where $\Theta(x)=1$ for $x>0$ and $\Theta(x)=0$ for $x<0$. In Fourier space $T_+$ is multiplication by $(\frac{1}{2}+ik)^{-1}$ and $T_-$ by $(\frac{1}{2}-ik)^{-1}$. Eigenfunctions of $T_+$, $T_-$ are the exponentials $\psi_a(x)=e^{-ax}$. For $a<\frac{1}{2}$ one has
\begin{equation}\label{4.11a}
T_+\psi_a(x)=\frac{1}{\frac{1}{2}-a}\psi_a(x)
\end{equation}
and for $a>-\frac{1}{2}$ one has
\begin{equation}\label{4.11b}
T_-\psi_a(x)=\frac{1}{\frac{1}{2}+a}\psi_a(x).
\end{equation}
For $|d|\leq m$ let, as an operator in $L^2(\R)$,
\begin{equation}\label{4.11c}
K_{m,d}= L P_- R
\end{equation}
with
\begin{equation}
L=T_+^{m+d} T_-^{-(m-d)}, \quad R=T_+^{-(m+d)} T_-^{m-d}.
\end{equation}
It follows that, for $a\in (-1/2,1/2)$,
\begin{eqnarray}\label{eqEV}
(R \psi_{-a})(x)&=&Z(a) \psi_{-a}(x), \\
(L^* \psi_a)(x)&=&Z(a)^{-1} \psi_a(x)\nonumber
\end{eqnarray}
where
\begin{equation}
Z(a)=\frac{(\frac12+a)^{m+d}}{(\frac12-a)^{m-d}}.
\end{equation}

For later use we provide a representation of the kernel of $K_{m,d}$. This kernel has a singular part, which is concentrated on the diagonal $\{x=y\}$. In the computations only the regular part will be used, hence only it is displayed. Since in Fourier space $T_+$, resp.\ $T_-$, is the operator of multiplication by $(\frac12+ik)^{-1}$, resp.\ $(\frac12-ik)^{-1}$, with the change of variable $\frac12-ik=z+\rho$ we obtain an integral expression for the regular part of $L$ and $R$. Let $\Gamma_p$ be a path around the pole $p$ oriented anti-clockwise. Then the regular part of the kernels are
\begin{equation}\label{eq3.19}
L(x,y)=\frac{1}{-2\pi i} e^{(1/2-\rho)(x-y)} I_{m,d}(x-y),\quad x-y>0,
\end{equation}
where
\begin{equation}\label{eq3.20}
I_{m,d}(x-y)=\int_{\Gamma_{1-\rho}}\dx z e^{-z (x-y)} \frac{(z+\rho)^{m-d}}{(1-\rho-z)^{m+d}},
\end{equation}
and similarly
\begin{equation}\label{eq3.21}
R(x,y)=\frac{1}{2\pi i} e^{(1/2-\rho)(x-y)} \tilde I_{m,d}(y-x),\quad x-y<0,
\end{equation}
where
\begin{equation}\label{eq3.22}
\tilde I_{m,d}(y-x)=\int_{\Gamma_{-\rho}}\dx z e^{z (y-x)} \frac{(1-\rho-z)^{m+d}}{(\rho+z)^{m-d}}.
\end{equation}
Let $P_u$ be the projection onto $[u,\infty)$. Then for any $u>0$ one has
\begin{equation}
\big(P_u K_{m,d} P_u\big)(x,y) = \Theta(x-u) \int_{\R_-}\dx w L(x,w) R(w,y) \Theta(y-u).
\end{equation}

As explained in Appendix~\ref{AppLaguerre}, the regular part of $K_{m,d}$ is a similarity transformed Laguerre kernel. Hence we refer to $K_{m,d}$ also as Laguerre kernel. It has the following properties, which are proved in Appendix~\ref{AppProp4}. One could also arrive at an equivalent kernel by taking the exponential limit of the geometric case studied by Okounkov in~\cite{Ok01}.

\begin{proposition}\label{PropInverse}
Let $u>0$. Then $\|P_u K_{m,d} P_u\|<1$. In addition, for $a>0$,
\begin{equation}\label{eq3.22b}
P_u(\Id-K_{m,d})\psi_a\in L^2(\R),\quad P_u(\Id-K_{m,d})^* \psi_a\in L^2(\R)
\end{equation}
with a norm uniformly bounded in $u$.
\end{proposition}

With these preparations we state the relation between the equal time kernel of (\ref{4.8}) and the Laguerre kernel.
\begin{proposition}\label{Prop4}
Let $0<a$, $b<\frac{1}{2}$. Then for $|d|<m$, $x,y>0$, one has the
identity
\begin{eqnarray}\label{4.13}
K^{(2m+1)}_{a,b}(2d,x;2d,y)&=& K^{a,b}_{m,d}(x,y)\\
&=&K_{m,d}(x,y)+\frac{1}{Z_{a,b}}(\Id-K_{m,d})\psi_b(x)(\Id-K_{m,d})^*
\psi_a(y)\nonumber
\end{eqnarray}
with
\begin{equation}\label{4.14}
Z_{a,b}=\frac{1}{a+b}\Big(\frac{1-2a}{1+2a}\Big)^m
\Big(\frac{1-2b}{1+2b}\Big)^m \Big(\frac{1}{4}-a^2\Big)^{-d}
\Big(\frac{1}{4}-b^2\Big)^{-d}.
\end{equation}
\end{proposition}

\section{Shift construction}\label{sect4}
Let us consider the Laguerre line ensemble with boundary values $a,b>0$ and denote
its weight by $\mathbbm{W}_{a,b}$. Under $\mathbbm{W}_{a,b}$ we want to study the
weight of $\{h_0(j,\tau)\leq u\}$ denoted by $\mathbbm{W}_{a,b} (\{h_0(j,\tau)\leq
u\})$. More general events could be investigated, but there is no need in our
context. We set $w_{a,b}(0,0)=v$ and recall that its weight is given by $e^{-v(a+b)}$, $v\geq 0$. We display the explicit dependence of $\mathbbm{W}_{a,b}$ on $v$ as
$\mathbbm{W}_{a,b}(\cdot,v)$.

From the construction of the Laguerre line ensemble one has, for $v>0$, $v+\delta>0$, the shift
\begin{equation}\label{5.1}
\mathbbm{W}_{a,b}(\{h_0(j,\tau)\leq u\},v+\delta)=
e^{-(a+b)\delta}\mathbbm{W}_{a,b}(\{h_0(j,\tau)+\delta\leq u\},v)
\end{equation}
and differentiating
\begin{eqnarray}\label{5.2}
\frac{\partial}{\partial v}\mathbbm{W}_{a,b}(\{h_0(j,\tau)\leq u\},v)&=&
-(a+b) \mathbbm{W}_{a,b}(\{h_0(j,\tau)\leq u\},v)\nonumber\\
& &- \frac{\partial}{\partial u}\mathbbm{W}_{a,b}(\{h_0(j,\tau)\leq u\},v).
\end{eqnarray}
Since $\mathbbm{W}_{a,b}(\cdot)=\int^\infty_0 \dx v \mathbbm{W}_{a,b}(\cdot,v)$, by
integrating in $v$,
\begin{eqnarray}\label{5.3}
\mathbbm{W}_{a,b}(\{h_0(j,\tau)\leq u\},0)&=&
\Dt{}{u}\mathbbm{W}_{a,b}(\{h_0(j,\tau)\leq u\})\nonumber\\
&&+(a+b) \mathbbm{W}_{a,b}(\{h_0(j,\tau)\leq u\}).
\end{eqnarray}
Note that the left hand side is the weight for $w_{a,b}(0,0)=0$.

Let $Z_{a,b}(v)= \mathbbm{W}_{a,b}(\{h_0(j,\tau)<\infty\},v)$ and
$Z_{a,b}=\int^\infty_0 \dx v Z_{a,b}(v)$. Then, taking $u\to\infty$ in (\ref{5.3}),
\begin{equation}\label{5.4}
Z_{a,b}(0)=(a+b)Z_{a,b},
\end{equation}
$Z_{a,b}$ given in (\ref{4.14}).

Let $\Pb_0^{a,b}$ be the probability for the Laguerre line ensemble in case $w_{a,b}(0,0)=0$ and $\Pb_{a,b}$ the one in case $w_{a,b}(0,0)$ is exponentially distributed with mean $(a+b)^{-1}$, as in Definition~\ref{defin3.1}. Then, by (\ref{5.3}) and (\ref{5.4}),
\begin{eqnarray}\label{5.5}
\Pb_0^{a,b} (\{h_0(j,\tau)\leq u\})&=&
\frac{1}{a+b}\Big(\Dt{}{u}\Pb_{a,b}(\{h_0(j,\tau)\leq u\})\nonumber\\
& &+(a+b) \Pb_{a,b}(\{h_0(j,\tau)\leq u\})\Big)
\end{eqnarray}
for $u>0$.

For determinantal processes probabilities as on the right hand side of (\ref{5.5}) are easily computed with the result
\begin{equation}\label{5.6}
\Pb_{a,b} (\{h_0(j,\tau)\leq u\})= \det(\Id-P_u K^{a,b}_{m,d} P_u),
\end{equation}
where, as before, $\tau=2m+1$, $j=2d$, and $P_u$ projects onto the interval $[u,\infty)$. The determinant is regarded in $L^2(\R)$ and the identity (\ref{5.6}) makes sense only for $u>0$. Thus we fix $u>0$ throughout. Since by (\ref{4.13}) $P_u K^{a,b}_{m,d} P_u$ is a rank one perturbation of $P_u K_{m,d} P_u$ and since $\Id-P_u K_{m,d} P_u$ is invertible, compare with Proposition~\ref{PropInverse}, one arrives at
\begin{equation}\label{5.7}
\det(\Id-P_u K^{a,b}_{m,d} P_u)= \det(\Id-P_u K_{m,d} P_u)(a+b) G^{a,b}(u)
\end{equation}
with
\begin{equation}\label{5.8}
(a+b) G^{a,b}(u)=1-\frac{1}{Z_{a,b}}\langle\psi_a,(\Id-K_{m,d})P_u(\Id-P_uK_{m,d}P_u)^{-1}
P_u(\Id-K_{m,d}) \psi_b\rangle
\end{equation}
with $\langle \cdot , \cdot \rangle$ denoting the inner product in $L^2(\R)$.

We also define
\begin{equation}\label{5.9}
F(u)= \det(\Id-P_u K_{m,d} P_u)
\end{equation}
and supply the $m,d$ dependence of $G^{a,b}(u)$ and of $F(u)$ when needed. We summarize as
\begin{proposition}\label{Prop5}
Let $w_{a,b}(i,j)$ be as in Definition~\ref{defin3.1}, except for $w_{a,b}(0,0)$ for which we set $w_{a,b}(0,0)=0$. Let $h_0$ be the corresponding top line as given in (\ref{4.3}), (\ref{4.4}). Then for $u>0$
\begin{equation}\label{5.10}
\Pb_0^{a,b} (\{h_0(j,\tau)\leq u\})= \Dt{}{u}\big(F(u)
G^{a,b}(u)\big) +F(u) (a+b) G^{a,b}(u),
\end{equation}
where $F(u)$ is given in (\ref{5.9}) and $G^{a,b}(u)$ in (\ref{5.8}).
\end{proposition}

\section{Analytic continuation}\label{sect5}
We have to extend the validity of (\ref{5.10}) from $0<a,b<1/2$ to $a+b=0$, which will be achieved by proving that both sides of (\ref{5.10}) are analytic.

\begin{proposition}\label{propAnalytic0}
The map $(a,b)\mapsto \Pb_0^{a,b}(\{h_0(j,\tau)\leq u\})$ is real analytic for $a,b>-1/2$.
\end{proposition}
\begin{proof}
$h_0(j,\tau)$ is measurable with respect to the $\sigma$-algebra generated by $w(i,0)$, $w(0,j)$, $i,j=1,\ldots,\tau$. Let
\begin{eqnarray}
& & V_u(\xi_1,\ldots,\xi_\tau,\eta_1,\ldots,\eta_\tau)\nonumber \\
& & = \Pb_0^{a,b}\big(\{h_0(j,\tau)\leq u\} | w(i,0)=\xi_i, w(0,j)=\eta_j, i,j=1,\ldots,\tau\big)
\end{eqnarray}
as conditional probability. Clearly $V_u$ does not depend on $a,b$ and $0\leq V_u\leq 1$. Then
\begin{eqnarray}
\Pb_0^{a,b}(\{h_0(j,\tau)\leq u\})&=& \int_{\R_+^{2\tau}}\prod_{k=1}^\tau \big(\dx \xi_k (\tfrac12+a)e^{-(1+2a)\xi_k/2}\big) \\
&\times &\prod_{k=1}^\tau \big(\dx \eta_k (\tfrac12+b)e^{-(1+2b)\eta_k/2}\big) V_{u}(\xi_1,\ldots,\xi_\tau,\eta_1,\ldots,\eta_\tau),\nonumber
\end{eqnarray}
which by inspection is real analytic for $a,b>-1/2$.
\end{proof}

\begin{proposition}\label{propAnalytic}
Let $u>0$ and let $G^{a,b}(u)$ be given by (\ref{5.8}). Then $(a,b)\mapsto G^{a,b}(u)$ extends to a real analytic function for $a,b\in (-1/2,1/2)$.
\end{proposition}
\begin{proof}
We repeat Eq.\ (\ref{5.8}),
\begin{equation}\label{eqGab}
(a+b) Z_{a,b} G^{a,b}(u)= Z_{a,b}-\langle \psi_a , (\Id-K_{m,d}) P_u (\Id-P_u K_{m,d} P_u)^{-1} P_u (\Id-K_{m,d}) \psi_b\rangle
\end{equation}
with $\psi_a(x)=e^{-a x}$.

First remark that $(a+b)Z_{a,b}$ is real analytic for $a,b\in (-1/2,1/2)$. On the other hand, in (\ref{eqGab}) the first (thus also the second) term diverges as $a+b\to 0$. Thus we have to find another representation of $G^{a,b}$ such that both terms remain finite in the $a+b\to 0$ limit. From each term we subtract the quantity $\langle \psi_a, P_u (\Id-K_{m,d}) \psi_b\rangle$ and obtain
\begin{eqnarray}\label{eqGab2}
\textrm{r.h.s.\ of }(\ref{eqGab}) &=& \big(Z_{a,b}-\langle \psi_a, P_u \psi_b\rangle\big) + \langle \psi_a, P_u K_{m,d}\psi_b\rangle \\
& &- \langle \psi_a , Q_u (\Id-K_{m,d}) P_u (\Id-P_u K_{m,d} P_u)^{-1} P_u (\Id-K_{m,d}) \psi_b\rangle \nonumber
\end{eqnarray}
where $Q_u=\Id-P_u$. With this rearrangement one singles out the divergence which now are in $Z_{a,b}$ and $\langle \psi_a, P_u \psi_b\rangle$ only. We discuss the analytic continuation from $a,b\in (0,1/2)$ to $a,b\in (-1/2,1/2)$ for the three terms separately, where we use the properties (see proof of Proposition~\ref{PropInverse})
\begin{eqnarray}\label{eq5.4}
|I_{m,d}(z)|&\leq& 2\pi C_{m,d} e^{-\beta_1 z}\textrm{ for any }0<\beta_1<1-\rho,\nonumber \\
|\tilde I_{m,d}(z)|&\leq& 2\pi \tilde C_{m,d} e^{-\beta_2 z}\textrm{ for any }0<\beta_2<\rho.
\end{eqnarray}

\vspace{6pt} \noindent {\it Term $Z_{a,b}-\langle \psi_a, P_u \psi_b\rangle$.}
Using the expression for $Z_{a,b}$, see (\ref{4.14}), we obtain
\begin{eqnarray}\label{eq5.7}
& &\langle \psi_a , (\Id-K) \psi_b\rangle-\langle\psi_a, P_u \psi_b\rangle \nonumber \\
&&= \frac{1}{a+b}\Big(\Big(\frac{1-2a}{1+2a}\frac{1-2b}{1+2b}\Big)^{m} \Big(\frac{1-4b^2}{1-4a^2}\Big)^{d}-e^{-(a+b) u}\Big).
\end{eqnarray}
(\ref{eq5.7}) is analytic for $a,b\in (-1/2,1/2)$ because the two terms in the bracket are $1+\Or(a+b)$ when $b+a\to 0$.

\vspace{6pt} \noindent {\it Term $\langle \psi_a, P_u K_{m,d}\psi_b\rangle$.}
Using (\ref{eqEV}) and (\ref{eq3.19}) one obtains
\begin{equation}\label{eq5.9}
\langle \psi_a, P_u K_{m,d}\psi_b\rangle = Z(-b)\int_u^\infty\dx x \int_{-\infty}^0\dx y \frac{I_{m,d}(x-y)}{-2\pi i} e^{-x(a+\rho-\frac12)} e^{y(\rho-\frac12-b)},
\end{equation}
where the function $I_{m,d}(z)$ is given in (\ref{eq3.20}). $Z(-b)$ is analytic if $b>-1/2$. Thus the integrand is bounded by $C_{m,d}e^{-x(\beta_1+a+\rho-\frac12)}e^{y(\beta_1-b+\rho-\frac12)}$. The condition $\beta_1<1-\rho$ implies that the integrand is exponentially decaying in $x-y$ provided that $a>-1/2$ and $b<1/2$. Thus r.h.s.\ of (\ref{eq5.9}) is real analytic for $a,b\in (-1/2,1/2)$.

\vspace{6pt} \noindent {\it Term $\langle \psi_a , Q_u (\Id-K_{m,d}) P_u (\Id-P_u K_{m,d} P_u)^{-1} P_u (\Id-K_{m,d}) \psi_b\rangle$.}
The object to consider is
\begin{equation}\label{eq5.7b}
\int_u^\infty\dx x \int_u^\infty \dx y f_a(x) (\Id-P_u K_{m,d} P_u)^{-1}(x,y) g_b(y)
\end{equation}
with $f_a(x)=-(K_{m,d}^* Q_u \psi_a)(x)$ and $g_b(y)=((\Id-K_{m,d})\psi_b)(y)$. Explicitly,
\begin{eqnarray}
f_a(x)&=&e^{-(\frac12-\rho)x}\int_u^\infty\dx w \int_{-\infty}^0\dx z e^{-w(a+\rho-\frac12)}\frac{I_{m,d}(w-z)}{-2\pi i} \frac{\tilde I_{m,d}(x-z)}{2\pi i} \nonumber \\
& & - Z(a)^{-1} e^{-(\frac12-\rho)x} \int_{-\infty}^0\dx z e^{-z(a+\rho-\frac12)} \frac{\tilde I_{m,d}(x-z)}{2\pi i}
\end{eqnarray}
and
\begin{equation}
g_b(x)=e^{-b x}-e^{-(\rho-\frac12)x}Z(-b)\int_{-\infty}^0\dx z \frac{I_{m,d}(x-z)}{-2\pi i}e^{z(\rho-\frac12-b)}.
\end{equation}
Using $(\Id-P_u K_{m,d} P_u)^{-1} = \Id + (\Id-P_u K_{m,d} P_u)^{-1}P_u K_{m,d} P_u$ we rewrite (\ref{eq5.7b}) as
\begin{equation}\label{eq5.10}
\langle f_a, P_u g_b\rangle + \langle f_a, P_u (\Id-P_u K_{m,d} P_u)^{-1} P_u \tilde g_b\rangle
\end{equation}
with $\tilde g_b= K_{m,d} P_u g_b$. Using (\ref{eq5.4}) one deduces that, for $a\in (-1/2,1/2)$,
\begin{equation}
f_a \in L^2((0,\infty),e^{\mu x}\dx x)\quad \textrm{for all }\mu<1/2
\end{equation}
and, for $b\in (-1/2,1/2)$,
\begin{equation}
g_b \in L^2((0,\infty),e^{-\mu x}\dx x)\quad \textrm{for all }\mu<-b<\frac12.
\end{equation}
This implies that
\begin{equation}
|\langle f_a, P_u g_b\rangle| \leq C \int_u^\infty \dx x e^{-(\beta_2+\frac12-\rho)x}(e^{-b x}+e^{-(\beta_1+\rho-\frac12)x})
\end{equation}
for some finite constant $C$. One has $\beta_2+\frac12-\rho+b\to \frac12+b>0$ as $\beta_2\to \rho$ and $\beta_1+\beta_2>0$, from which it follows that $\langle f_a, P_u g_b\rangle$ is real analytic for $a,b\in (-1/2,1/2)$.

For the other term in (\ref{eq5.10}) we have to compute $\tilde g_b(x)$. We obtain
\begin{equation}
\tilde g_b(x) = e^{-(\rho-\frac12)x}\int_u^\infty\dx w \int_{-\infty}^0\dx z \frac{I_{m,d}(x-z)}{-2\pi i} \frac{I_{m,d}(w-z)}{2\pi i} e^{-w(\frac12-\rho)} g_b(w).
\end{equation}
Then
\begin{equation}
|\tilde g_b(x)| \leq C e^{-(\beta_1+\rho-\frac12)x}\int_u^\infty \dx w \int_{-\infty}^0\dx z e^{(\beta_1+\beta_2)z} (e^{-(b+\beta_2+\frac12-\rho)w}+e^{-(\beta_1+\beta_2)})
\end{equation}
for some constant $C$. One has $\beta_1+\rho-\frac12\to \frac12$ as $\beta_1\to 1-\rho$, $\beta_2+\frac12-\rho-b\to \frac12-b>0$ as $\beta_2\to \rho$, and $\beta_1+\beta_2$. It follows
\begin{equation}
\tilde g_b \in L^2((0,\infty),e^{\mu x}\dx x)\quad \textrm{for all }\mu<1/2.
\end{equation}
From this one deduces that $\langle f_a, P_u (\Id-P_u K_{m,d} P_u)^{-1} P_u \tilde g_b\rangle$ is also real analytic for $a,b\in (-1/2,1/2)$.

\vspace{6pt}\noindent {\it A second representation.} Another way of eliminating the divergence consists in subtracting the quantity $\langle \psi_a, (\Id-K_{m,d}) P_u\psi_b\rangle$ from each term. This leads to
\begin{eqnarray}\label{eqGab2bis}
\textrm{r.h.s.\ of }(\ref{eqGab}) &=& \big(Z_{a,b}-\langle \psi_a, P_u \psi_b\rangle\big) + \langle \psi_a, K_{m,d} P_u\psi_b\rangle \\
& &- \langle \psi_a , (\Id-K_{m,d}) P_u (\Id-P_u K_{m,d} P_u)^{-1} P_u (\Id-K_{m,d}) Q_u \psi_b\rangle. \nonumber
\end{eqnarray}
By the same argument as above one shows that the terms in this second representations are analytic for $a,b\in (-1/2,1/2)$.
\end{proof}

By Propositons~\ref{propAnalytic0} and~\ref{propAnalytic} one can take the limit $b\to -a$ without loosing the identity (\ref{5.10}). We remark that $\lim_{b\to -a} (a+b) Z_{a,b}=1$. In addition, by (\ref{eq5.7}),
\begin{equation}
\lim_{b\to -a}\langle \psi_a,(\Id-K_{m,d})\psi_b\rangle-\langle\psi_a, P_u \psi_b\rangle = u+\frac{2ad-m}{\frac14-a^2}.
\end{equation}
Let us denote the limit
\begin{equation}
\lim_{b\to -a} G^{a,b}(u)=G_0(u),
\end{equation}
with the $a$-dependence understood implicitly. Then by (\ref{eqGab2})
\begin{eqnarray}\label{eqGabPos}
G_0(u)&=&\lim_{b\to -a} G^{a,b}(u) = \Big(u+\frac{2ad-m}{\frac14-a^2}\Big)+\langle \psi_a,P_u K_{m,d}\psi_{-a}\rangle  \\
& &- \langle \psi_a , Q_u (\Id-K_{m,d}) P_u (\Id-P_u K_{m,d} P_u)^{-1} P_u (\Id-K_{m,d}) \psi_{-a}\rangle.\nonumber
\end{eqnarray}
Alternatively, using (\ref{eqGab2bis}) one can write
\begin{eqnarray}\label{eqGabNeg}
G_0(u)&=&\lim_{b\to -a} G^{a,b}(u) = \Big(u+\frac{2ad-m}{\frac14-a^2}\Big)+\langle \psi_a, K_{m,d} P_u\psi_{-a}\rangle  \\
& &- \langle \psi_a , (\Id-K_{m,d}) P_u (\Id-P_u K_{m,d} P_u)^{-1} P_u (\Id-K_{m,d}) Q_u\psi_{-a}\rangle.\nonumber
\end{eqnarray}

With the same convention, let us denote
\begin{equation}
\lim_{b\to -a}\Pb_0^{a,b}=\Pb_0.
\end{equation}
Recall that by construction $\Pb_0$ is the probability measure for the family $w(i,j)$, $i,j\in\N$, of independent exponentially distributed random variables such that $w(0,0)=0$, $w(i,0)$ has mean $(1-\rho)^{-1}$ for $i\geq 1$, $w(0,j)$ has mean $\rho^{-1}$ for $j\geq 1$, and otherwise $w(i,j)$ has mean $1$. This is precisely the $w$-marginal of $\Q$ in Section~\ref{sect2}. Also, by definition, $h^0(j,\tau)=h_0(j,\tau)$ pathwise. Thus one concludes
\begin{proposition}\label{prop5.3} For $0< c_1 < c_2 < \infty$ it holds
\begin{eqnarray}
\int_{c_1}^{c_2} \dx u \Q\big(\{h^0(j,\tau)\leq u\}\big) &=& \int_{c_1}^{c_2} \dx u \Pb_0\big(\{h_0(j,\tau)\leq u\}\big)  \nonumber\\
&=& F(c_2) G_0(c_2)-F(c_1) G_0(c_1).
\end{eqnarray}
\end{proposition}

To prove Theorem~\ref{Thm2} one has to investigate the asymptotics of $F(u)G_0(u)$ under the scaling (\ref{3.12}).

\section{Edge scaling}\label{sectEdge}
Following (\ref{3.12}) we set
\begin{eqnarray}\label{eqScalingt}
2m = \tau -1&=& \lfloor (1-2\chi) t + 2 w (1-2\rho) \chi^{1/3} t^{2/3}\rfloor,\nonumber \\
2d = j &=& \lfloor (1-2\rho) t +2 w \chi^{1/3} t^{2/3}\rfloor, \\
u&=& \lfloor t+s\chi^{-1/3} t^{1/3}\rfloor ,\nonumber
\end{eqnarray}
with $\chi=\rho(1-\rho)$. Then, by Proposition~\ref{prop5.3}, the proof of the limit (\ref{3.14}) in Theorem~\ref{Thm2} reduces to the large $t$ limit of
\begin{equation}
\chi^{-1/3} t^{1/3} G_0(t+s\chi^{-1/3} t^{1/3})
\end{equation}
where the prefactor takes into account the scaling of $\Dt{}{u}$,
as well as the large $t$ limit of the Fredholm determinant
\begin{equation}
\det(\Id-P_u K_{m,d}P_u)
\end{equation}
on $L^2(\R_+)$. This latter limit has been studied by Johansson, see Theorem 1.6 in \cite{Jo00b}.
\begin{theorem}\label{thmJo}{\rm (Johansson)} Let $u=t+s\chi^{-1/3} t^{1/3}$ and $m,d$ as in (\ref{eqScalingt}). Then
\begin{equation}
\lim_{t\to\infty} \det(\Id-P_u K_{m,d}P_u) = F_{\rm GUE}(s+w^2),
\end{equation}
where $F_{\rm GUE}$ is the GUE Tracy-Widom distribution function~\cite{TW94}.
\end{theorem}
In order to state the limit of $G_0$ we have to introduce some auxiliary quantities. We define the functions
\begin{equation}
\varphi_{w,s}(z)=\Ai(w^2+s+z) e^{w(w^2+s+z)} e^{-\frac13 w^3}
\end{equation}
and
\begin{eqnarray}
S_{w,s} &=& s + \int_{\R_+^2} \dx x \dx y \varphi_{w,s}(x+y), \nonumber \\
\Phi_{w,s}(\xi)&=& e^{w \xi} \int_{\R_+}\dx y \varphi_{-w,s}(y+\xi) \Big(1-\int_{\R_+}\dx x \varphi_{w,s}(x+y)\Big),\\
\Psi_{w,s}(\xi)&=& e^{-w \xi} \Big(1-\int_{\R_+}\dx x \varphi_{w,s}(x+\xi)\Big). \nonumber
\end{eqnarray}
These functions can be written as a single integral using the identity (\ref{Airy3}), i.e.,
\begin{equation}
1-\int_{\R_+}\dx x \varphi_{w,s}(x+y) = \int_{\R_-}\dx x \varphi_{w,s}(x+y)
\end{equation}
for $w>0$ (for $w=0$ the same holds but only as a improper Riemann integral). Using a contour integral representation, in case $w >0$, one rewrites $S_{w,s}$ as (see Section~\ref{Sws})
\begin{equation}\label{eqSbis}
S_{w,s} = \int_{\R_-^2} \dx x \dx y \varphi_{w,s}(x+y).
\end{equation}
It is easy to see, using the super-exponential decay of the Airy function, that $\Phi_{w,s}\in L^2(\R_+)$ for $w\geq 0$, $\Psi_{w,s}\in L^2(\R_+)$ for $w>0$, and in the case $w=0$, $\Psi_{0,s}-1\in L^2(\R_+)$.
Finally we denote by $K_{\Ai,q}$ the operator with kernel
\begin{equation}
K_{\Ai,q} (\xi_1,\xi_2)= \int_{\R_+} \dx z \Ai(q+z+\xi_1)\Ai(q+z+\xi_2).
\end{equation}
$K_{\Ai,q}$ is the Airy kernel shifted by $q$.

\begin{theorem}\label{thmG0}
\begin{eqnarray}\label{eqFinal}
& & \lim_{t\to\infty} \chi^{1/3} t^{-1/3} G_0(t+s\chi^{-1/3} t^{1/3}) \\
&&= S_{|w|,s} + \int_{\R_+} \dx z \Phi_{|w|,s}(z)P_0 (\Id-P_0K_{\Ai,w^2+s}P_0)^{-1}P_0\Psi_{|w|,s}(z) \nonumber \\
& & = g(s+w^2,w).\nonumber
\end{eqnarray}
\end{theorem}
Theorem~\ref{thmJo} and Theorem~\ref{thmG0}, in conjunction with Proposition~\ref{prop5.3}, furnish the proof of Theorem~\ref{Thm2} and hence of our main result Theorem~\ref{MainThm}. Beyond the existence of the limit, they also implies Corollary~\ref{corollario}.

\vspace{6pt}
\begin{proofOF}{Theorem~\ref{thmG0}}\\
\textbf{A change of variable: from $t$ to $N$.}
We change variables with the effect to have $N$ instead of $t$ as large parameter. This is not really necessary but simplifies our computations. Let us define $N=m-d=(\tau-1-j)/2$ and $N+\alpha=m+d$. The relevant parameters for what follows are $\alpha=2d=j$ and $u$ which, in terms of $N$, are given by
\begin{eqnarray}\label{eqScalingN}
\alpha &=& \frac{1-2\rho}{\rho^2}N + 2 w \frac{(1-\rho)^{4/3}}{\rho^2} N^{2/3}\nonumber \\ & &+ \Big(\frac83 w^2(1-\rho)+s(1-2\rho)\Big)\frac{(1-\rho)^{2/3}}{\rho^2} N^{1/3} +\Or(1),\\
u &=& \frac{N}{\rho^2} + 2 w \frac{(1-\rho)^{1/3}}{\rho^2} N^{2/3}+
\Big(\frac83 w^2+s\Big)\frac{(1-\rho)^{2/3}}{\rho^2} N^{1/3} +\Or(1).\nonumber
\end{eqnarray}
Moreover the scaling $\chi^{-1/3}t^{1/3}$ writes as
\begin{equation}
\chi^{-1/3}t^{1/3} = \kappa N^{1/3}+\Or(1)
\end{equation}
with
\begin{equation}
\kappa=\rho^{-1} (1-\rho)^{-1/3}.
\end{equation}

After edge scaling the terms of $G_0$ will be expressed via the functions $H_N$ and $\tilde H_N$, defined as
\begin{eqnarray}
H_N(y) &=& Z(a) \frac{\kappa N^{1/3}}{-2\pi i}I_{N+\alpha/2,\alpha/2}(u+y \kappa N^{1/3}) ,\\
\tilde H_N(y) &=& Z(a)^{-1} \frac{\kappa N^{1/3}}{2\pi i}\tilde I_{N+\alpha/2,\alpha/2}(u+y \kappa N^{1/3}). \nonumber
\end{eqnarray}
Using the bounds on $I$ and $\tilde I$ of Section~\ref{sect6} we obtain, for any $\beta>0$ fixed,
\begin{eqnarray}\label{Bound1}
|H_N(y)|&\leq & C_\beta e^{-\beta y}, \\
|\tilde H_N(y)|&\leq & C_\beta e^{-\beta y}\nonumber
\end{eqnarray}
for some $C_\beta>0$ independent of $N$ and $y\geq 0$. Moreover we also have the pointwise convergence
\begin{eqnarray}\label{Asympt7}
\lim_{N\to\infty} H_N(y) &=& \Ai(w^2+s+y) e^{w(w^2+s+y)} e^{-\frac13 w^3} = \varphi_{w,s}(y), \\
\lim_{N\to\infty} \tilde H_N(y) &=& \Ai(w^2+s+y) e^{-w(w^2+s+y)} e^{\frac13 w^3}=\varphi_{-w,s}(y). \nonumber
\end{eqnarray}
We simplify the notations in this proof by setting $K=K_{N+\alpha/2,\alpha/2}$.

\vspace{6pt}
\noindent \textbf{The estimate of the terms for $\kappa^{-1} N^{-1/3} G_0(u+s\kappa N^{1/3})$}.\\[6pt]
\noindent \textit{First term of (\ref{eqGabPos}) and (\ref{eqGabNeg}).} Using (\ref{eqScalingN}), $2d=\alpha$, and $m-d=N$, we have
\begin{equation}\label{eqFirst}
\lim_{N\to\infty}u+\frac{2ad-m}{\frac14-a^2} = s.
\end{equation}

\vspace{6pt}
\noindent \textit{Second term of (\ref{eqGabPos}) for $w \geq 0$.} We compute the limit
\begin{equation}
\lim_{N\to\infty} \kappa^{-1} N^{-1/3} \langle \psi_a,P_u K\psi_{-a}\rangle.
\end{equation}
$\psi_{-a}$ is eigenfunction of $R$, see (\ref{eqEV}). Thus we have
\begin{equation}
\langle\psi_a, P_u K \psi_{-a}\rangle = Z(a) \langle\psi_a, P_u L P_- \psi_{-a}\rangle = \kappa N^{1/3} \int_0^\infty \dx x \int_0^\infty \dx y H_N(x+y).
\end{equation}
Using the bound (\ref{Bound1}) we can apply dominated convergence. Then by the pointwise limit (\ref{Asympt7}) we have
\begin{equation}\label{eqSecond}
\lim_{N\to\infty} \kappa^{-1} N^{-1/3} \langle \psi_a, P_u K \psi_{-a}\rangle =  \int_0^\infty \dx x \int_0^\infty \dx y \varphi_{w,s}(x+y).
\end{equation}

\vspace{6pt}
\noindent \textit{Second term of (\ref{eqGabNeg}) for $w \leq 0$.} This case is analogous to the previous one. One obtains
\begin{equation}\label{eqSecondB}
\lim_{N\to\infty} \kappa^{-1} N^{-1/3} \langle \psi_a, K  P_u\psi_{-a}\rangle = \int_0^\infty \dx x \int_0^\infty \dx y \varphi_{-w,s}(x+y).
\end{equation}
The sum of (\ref{eqFirst}) and (\ref{eqSecond}), resp.\ (\ref{eqSecondB}), yields $S_{|w|,s}$ as the first term in Theorem~\ref{thmG0}.

\vspace{6pt}
\noindent \textit{Third term of (\ref{eqGabPos}) for $w\geq 0$.}
The third term of $\kappa^{-1} N^{-1/3}G_0(u)$, including the prefactor $-1$, is
\begin{equation}\label{eq1}
\kappa^{-1} N^{-1/3} \langle \tilde \Phi_N, \tilde A_N \tilde \Psi_N\rangle
\end{equation}
with
\begin{eqnarray}
\tilde \Phi_N(x)&=&\big(K^*Q_u \psi_a\big)(x), \nonumber \\
\tilde \Psi_N(y)&=&\big((\Id-K)\psi_{-a}\big)(y), \\
\tilde A_N(x,y)&=& \big(P_u (\Id-P_u K P_u)^{-1} P_u\big)(x,y).\nonumber
\end{eqnarray}
To establish the scaling limit, $x=u+\xi \kappa N^{1/3}$, we define the rescaled quantities
\begin{eqnarray}\label{eq3}
\psi_a^{\ell}(\xi)&=&\psi_a(u+\xi\kappa N^{1/3}) \nu(\xi)^{-1}, \nonumber\\
\psi_{-a}^r(\xi)&=&\psi_{-a}(u+\xi\kappa N^{1/3}) \nu(\xi), \\
K_N^r(\xi_1,\xi_2)&=&\kappa N^{1/3} K(u+\xi_1\kappa N^{1/3},u+\xi_2\kappa N^{1/3})  \nu(\xi_1) \nu(\xi_2)^{-1}, \nonumber
\end{eqnarray}
where
\begin{equation}
\nu(\xi)=e^{-a (u+\xi \kappa N^{1/3})} e^{-w \xi}.
\end{equation}
Then (\ref{eq1}) becomes
\begin{equation}\label{eq2}
\langle \Phi_N, A_N \Psi_N\rangle
\end{equation}
with
\begin{eqnarray}\label{eqFct0}
\Phi_N(\xi_1)&=&\big(K_N^{r,*} Q_0 \psi_a^{\ell}\big)(\xi_1), \nonumber \\
\Psi_N(\xi_2)&=&\big((\Id-K_N^r)\psi_{-a}^r\big)(\xi_2), \\
A_N(\xi_1,\xi_2)&=& \big(P_0 (\Id-P_0 K_N^r P_0)^{-1} P_0\big)(\xi_1,\xi_2). \nonumber
\end{eqnarray}
We can rewrite the $\Phi_N$, $\Psi_N$, and $K_N^r$ by using the functions $H_N$ and $\tilde H_N$,
\begin{eqnarray}\label{eqFct}
\Phi_N(\xi_1) &=& e^{w \xi_1} \Big(\int_{\R_+} \dx y \tilde H_N(y+\xi_1)-\int_{\R_+^2}\dx x \dx y H_N(x+y) \tilde H_N(y+\xi_1)\Big), \nonumber \\
\Psi_N(\xi_2) &=& e^{-w \xi_2} \Big(1-\int_{\R_+}\dx y H_N(y+\xi_2)\Big), \\
K_N^r(\xi_1,\xi_2) &=& e^{-w\xi_1} e^{w\xi_2} \int_{\R_+} \dx x H_N(x+\xi_1) \tilde H_N(x+\xi_2). \nonumber
\end{eqnarray}

We want to avoid to write always the projection $P_0$. Therefore from now on $\langle \cdot , \cdot \rangle$ refers to the scalar product in $L^2(\R_+,\dx x)$, $\|\cdot \|$ is the corresponding norm, and the integral operators act in $L^2(\R_+,\dx x)$. First let us consider $w>0$. Let us denote $A=(\Id-K_{\Ai,w^2+s})^{-1}$. Then for finite $N$ we have the bound
\begin{eqnarray}\label{eq45}
& & |\langle \Phi_N, A_N \Psi_N\rangle-\langle \Phi_{w,s}, A\Psi_{w,s} \rangle| \leq \|\Phi_N\| \,\|A_N-A\|\, \|\Psi_N\| \\ & & + \|\Phi_N-\Phi_{w,s}\| \,\|A\|\, \|\Psi_N\|+\|\Phi_{w,s}\|\,\|A\|\,\|\Psi_N-\Psi_{w,s}\|.\nonumber
\end{eqnarray}
In Lemma~\ref{lemConv} we will prove that $\Phi_N$ converges to $\Phi_{w,s}$, $\Psi_N$ converges to $\Psi_{w,s}$, and $A_N$ converges to $(\Id-K_{\Ai,w^2+s})^{-1}$ in operator norm (in $L^2(\R_+)$ according to our convention). This implies
\begin{equation}
\lim_{N\to\infty}\langle \Phi_N, A_N \Psi_N\rangle = \langle \Phi_{w,s}, (\Id-K_{\Ai,w^2+s})^{-1}\Psi_{w,s} \rangle
\end{equation}
which is precisely the last term in (\ref{eqFinal}).

For the case $w=0$ we have to modify slightly the argument. In this case $\Psi_N$ is not in $L^2(\R_+)$, but $\Psi_N(\xi)-1=-\int_{\R_+}\dx y H_N(y+\xi)\in L^2(\R_+)$ and converges to $-\int_{\R_+} \dx x \varphi_{0,s}(x+\xi)$ in the $N\to\infty$ limit. We write
\begin{equation}\label{eq46}
\langle \Phi_N, A_N \Psi_N\rangle = \langle \Phi_N, A_N (\Psi_N-1)\rangle + \langle \Phi_N,A_N 1\rangle.
\end{equation}
Here $1$ denotes the constant function $1(x)=1$, for all $x\in\R$. By the same argument as for $w>0$, the first term in (\ref{eq46}) converges as
\begin{equation}
\lim_{N\to\infty}\langle \Phi_N, A_N (\Psi_N-1)\rangle= \langle \Phi_{w,s}, (\Id-K_{\Ai,w^2+s})^{-1}(\Psi_{w,s}-1) \rangle.
\end{equation}
The second term in (\ref{eq46}) can be rewritten as
\begin{equation}
\langle \Phi_N,A_N 1\rangle = \int_{\R_+^2} \dx \xi_1\dx \xi_2 \Phi_N(\xi_1)A_N(\xi_1,\xi_2).
\end{equation}
In Lemma~\ref{lemConv2} we will prove the convergence
\begin{equation}
\lim_{N\to\infty}\langle \Phi_N,A_N 1\rangle= \langle \Phi_{w,s}, (\Id-K_{\Ai,w^2+s})^{-1}1 \rangle.
\end{equation}

\vspace{6pt}
\noindent \textit{Third term of (\ref{eqGabNeg}) for $w\leq 0$.} The computations for this case are as before and the third term of  (\ref{eqGabNeg}) converges to
\begin{equation}
\langle \Psi_{-w,s}, (\Id-K_{\Ai,w^2+s})^{-1}\Phi_{-w,s} \rangle.
\end{equation}
Since $K_{\Ai,w^2+s}$ is symmetric, this concludes the proof of Theorem~\ref{thmG0}.
\end{proofOF}

\begin{lemma}\label{lemConv}
Let $w\geq 0$. Then
\begin{eqnarray}
& &\lim_{N\to\infty}\|\Phi_N-\Phi_{w,s}\| =0,\nonumber \\
& & \lim_{N\to\infty}\|\Psi_N-\Psi_{w,s}\| =0,\\
& & \lim_{N\to\infty} \|A_N-(\Id-K_{\Ai,w^2+s})^{-1}\| =0, \nonumber
\end{eqnarray}
where the functions $\Phi_N,\Psi_N$, and the integral kernel $A_N$ are defined in (\ref{eqFct0}) and (\ref{eqFct}).
\end{lemma}

\begin{proof} \textit{Convergence of $\Phi_N$.} Let us consider, for any fixed $\xi \geq 0$, the function $\Phi_N(\xi)$ defined in (\ref{eqFct}). We first show that
\begin{equation}
\lim_{N\to\infty} \Phi_N(\xi) = \Phi_{w,s}(\xi)
\end{equation}
pointwise. Using the exponential decay (\ref{Bound1}) of $H_N$ and $\tilde H_N$, we apply dominated convergence and exchange the integrals with the $N\to\infty$ limit. Then using the pointwise limit of $H_N$, see (\ref{Asympt7}), one obtains
\begin{equation}\label{eq54}
\lim_{N\to\infty} \Phi_N(\xi) = e^{w\xi} \int_{\R_+} \dx y \varphi_{-w,s}(y+\xi)- e^{w\xi} \int_{\R_+^2}\dx y \dx x \varphi_{-w,s}(y+\xi) \varphi_{w,s}(x+y),
\end{equation}
which is precisely $\Phi_{w,s}(\xi)$. By the exponential decay (\ref{Bound1}) with $\beta>w$ it follows that $\Phi_N \in L^2(\R_+)$. Moreover, $\Phi_N$ is uniformly bounded by an integrable function since, for all $\beta>w$,
\begin{equation}
|\Phi_N(\xi)| \leq C_\beta^2 e^{-\beta x}e^{-2\beta y} e^{-(\beta-w)\xi}.
\end{equation}
Therefore, by dominated convergence and by pointwise convergence (\ref{eq54}), we obtain
\begin{equation}
\lim_{N\to\infty} \| \Phi_N-\Phi_{w,s}\|^2 = \int_{\R_+} \dx \xi \lim_{N\to\infty} |\Phi_N(\xi)-\Phi_{w,s}(\xi)|^2 =0.
\end{equation}

\vspace{6pt}
\noindent\textit{Convergence of $\Psi_N$.} Let us consider, for $\xi \geq 0$, the function $\Psi_N(\xi)$ defined in (\ref{eqFct}). We first show that
\begin{equation}
\lim_{N\to\infty} \Psi_N(\xi) = \Psi_{w,s}(\xi)
\end{equation}
pointwise. As before, (\ref{Bound1}) allows us to exchange the limit and the integral with the result
\begin{equation}\label{eq5}
\lim_{N\to\infty} \Psi_N(\xi) = e^{-w \xi} \Big(1-\int_0^\infty\dx y \varphi_{w,s}(y+\xi)\Big).
\end{equation}
Then by (\ref{Bound1}), with $\beta>w$, $\Psi_N \in L^2(\R_+)$ and $\Psi_N$ is bounded by an integrable function. Therefore
\begin{equation}
\lim_{N\to\infty} \| \Psi_N-\Psi_{w,s}\|^2 = \int_{\R_+} \dx \xi \lim_{N\to\infty} |\Psi_N(\xi)-\Psi_{w,s}(\xi)|^2 =0.
\end{equation}

\vspace{6pt}
\noindent\textit{Convergence of $A_N$.} Denote $q=w^2+s$. We want to show that
\begin{equation}\label{eq16}
\lim_{N\to\infty} \|(\Id-K_N^r)^{-1}-(\Id-K_{\Ai,q})^{-1}\|=0
\end{equation}
in operator norm. Assume that we can show:\\[6pt]
1) $\lim_{N\to\infty}\|K_N^r-K_{\Ai,q}\|=0$,\\[6pt]
2) $\|(\Id-K_{\Ai,q})^{-1}\|< \infty$.\\[6pt]
Then with the notation $K_\e=K_{\Ai,q}-K_N^r$ we have
\begin{eqnarray}
& &\|(\Id-K_N^r)^{-1}-(\Id-K_{\Ai})^{-1}\| \nonumber \\
&&= \|[(\Id+(\Id-K_{\Ai,q})^{-1} K_\e)^{-1}-1] (\Id-K_{\Ai,q})^{-1}\| \nonumber \\
& &\leq \|(\Id-K_{\Ai,q})^{-1}\| \sum_{n\geq 1} \|(\Id-K_{\Ai,q})^{-1}\|^n \|K_\e\|^n
\end{eqnarray}
which converges to $0$ as $N\to\infty$. So we have to establish properties 1) and 2). Let us start with 1). For fixed $x,y\geq 0$, using (\ref{Bound1}) with $\beta>w$,  $K_N^r(x,y)$ is uniformly bounded by a function which is integrable in $\R_+^2$. Therefore, by dominated convergence and the pointwise limit
\begin{equation}
\lim_{N\to\infty}H_N(x+z) \tilde H_N(y+z)=\Ai(q+x+z)\Ai(q+y+z),
\end{equation}
one obtains
\begin{eqnarray}
& &\lim_{N\to\infty}|K_N^r(x,y)-K_{\Ai,q}(x,y)|\\
& & \leq \int_0^\infty\dx z [\lim_{N\to\infty}H_N(x+z) \tilde H_N(y+z)-\Ai(q+x+z)\Ai(q+y+z)]=0.\nonumber
\end{eqnarray}
Moreover, it is easy to see that $K_N^r$ is a Hilbert-Schmidt operator with norm uniformly bounded in $N$, and so is $K_{\Ai,q}$. Therefore
\begin{eqnarray}
& &\lim_{N\to\infty} \|K_N^r(x,y)-K_{\Ai,q}(x,y)\|^2 \leq \lim_{N\to\infty} \|K_N^r(x,y)-K_{\Ai,q}(x,y)\|^2_{\rm HS} \nonumber \\
& & =  \int_{\R_+^2}\dx x\dx y \lim_{N\to\infty}| K_N^r(x,y)-K_{\Ai,q}(x,y)|^2=0.
\end{eqnarray}
Next consider point 2). $\Id-K_{\Ai,q}$ is invertible as bounded operator, since for every $q\in \R$, $\|K_{\Ai,q}\|<1$. To establish the claim, let us denote by $K_{\Ai}$ the standard Airy operator with Airy kernel $K_{\Ai,0}$. Then
\begin{equation}
\|K_{\Ai,q}\|_{L^2(\R_+)}=\|K_{\Ai}\|_{L^2([q,\infty))}.
\end{equation}
$K_{\Ai}$ is an operator on $L^2([q,\infty))$ which is Hilbert-Schmidt. Therefore the norm of $K_{\Ai,q}$ on $L^2([q,\infty))$ equals its largest eigenvalue, $\lambda_0(q)$. In~\cite{TW94} it is shown that $\lambda_0(q)$ is monotonically decreasing in $q$ and is strictly less than $1$ for any $q>-\infty$ (it converges to $1$ as $q\to-\infty$).
\end{proof}

\begin{lemma}\label{lemConv2}
Let $w=0$, then
\begin{equation}\label{eq66}
\lim_{N\to\infty} \langle\Phi_N,A_N 1\rangle = \langle \Phi_{0,s}, (\Id-K_{\Ai,s})^{-1} 1\rangle.
\end{equation}
\end{lemma}
\begin{proof}
We first rewrite
\begin{equation}
\langle\Phi_N,A_N 1\rangle = \langle \Phi_N, 1\rangle+ \langle \Phi_N, (\Id-K_N^r)^{-1} K_N^r 1\rangle.
\end{equation}
The first term, $\langle \Phi_N, 1\rangle=\int_{\R_+}\dx x \Phi_N(x)$, converges to $\int_{\R_+}\dx x \Phi_{0,s}(x)=\langle \Phi_{0,s}, 1\rangle$ as $N\to\infty$ because $\Phi_N\in L^1(\R_+)$ with norm $L^1(\R_+)$ uniformly bounded in $N$. For the second term, define $\widehat \Phi_N = (\Id-K_N^{r,*})^{-1} \Phi_N$. From Lemma~\ref{lemConv} it follows that, for $N$ large enough, $\widehat \Phi_N \in L^2(\R_+)$ with uniformly bounded norm.  Then
\begin{equation}
\langle \Phi_N, (\Id-K_N^r)^{-1} K_N^r 1\rangle = \int_{\R_+}\dx x \int_{\R_+} \dx y \widehat \Phi_N(y) K_N^r(y,x) \equiv \int_{\R_+}\dx x D_N(x).
\end{equation}
$D_N(x)$ is bounded by an $L^1$ function independent of $N$. In fact, using the representation (\ref{eqFct}) of $K_N^r$, one writes
\begin{eqnarray}
|D_N(x)|&\leq& \int_{R_+}\dx \lambda |\tilde H_N(x+\lambda)| \int_{\R_+}\dx y |H_N(y+\lambda)|\, |\widehat \Phi_N(y)| \\
&\leq & \int_{R_+}\dx \lambda |\tilde H_N(x+\lambda)| \Big(\int_{\R_+}\dx y |H_N(y+\lambda)|^2\Big)^{1/2} \|\widehat \Phi_N\|_2\nonumber.
\end{eqnarray}
Since $H_N$ and $\tilde H_N$ decay exponentially, see (\ref{Bound1}), we have $|D_N(x)|\leq C e^{-x}$ for some $C>0$ independent of $N$. Therefore by dominated convergence we obtain
\begin{equation}
\lim_{N\to\infty} \langle \Phi_N, (\Id-K_N^r)^{-1} K_N^r 1\rangle = \int_{\R_+}\dx x \lim_{N\to\infty}\int_{\R_+}\dx y \widehat \Phi_N(y) K_N^r(y,x).
\end{equation}
This last integral can be interpreted as the $L^2(\R_+)$ scalar product between $\widehat \Phi_N$ and $K_N^r(\cdot,x)$. By Lemma~\ref{lemConv}, $\widehat \Phi_N$ converges to $(\Id-K_{\Ai,s})^{-1} \Phi_{0,s}$. Moreover, $K_N^r(\cdot,x)$ converges to $K_{\Ai,s}(\cdot,x)$, thus
\begin{eqnarray}
& &\lim_{N\to\infty} \langle \Phi_N, (\Id-K_N^r)^{-1} K_N^r 1\rangle \nonumber \\
&&= \int_{\R_+}\dx x \int_{\R_+}\dx y (\Id-K_{\Ai,s})^{-1} \Phi_{0,s}(y) K_{\Ai,s}(y,x) \\
&&=\langle \Phi_{0,s}, (\Id-K_{\Ai,s})^{-1} K_{\Ai,s} 1\rangle. \nonumber
\end{eqnarray}
Adding the first and second term one obtains (\ref{eq66}).
\end{proof}

\section{Asymptotics used in Section~\ref{sectEdge}}\label{sect6}
In this section $\alpha$ and $u$ are defined as in (\ref{eqScalingN}). First we summarize the asymptotic results required. Let $m=N+\alpha/2$ and $d=\alpha/2$.

\subsection{Asymptotics of $I_{m,d}(u+y \kappa N^{1/3})$, $0<\rho<1$}\label{AS}
{\bf \ref{AS}.1} For fixed $y\in \R$,
\begin{equation}\label{Asympt1}
Z(a) \kappa N^{1/3} I_{m,d}(u+y \kappa N^{1/3}) = -2\pi i \Ai(w^2+s+y) e^{w(w^2+s+y)} e^{-\frac13 w^3} +\Or(N^{-1/3}).
\end{equation}

\noindent {\bf \ref{AS}.2} For $0\ll L \leq y \leq \e N^{2/3}$, $\e>0$ small enough, $L>0$ large enough,
\begin{equation}\label{Asympt2}
|Z(a) \kappa N^{1/3} I_{m,d}(u+y \kappa N^{1/3})| \leq  C e^{-\frac13  y^{3/2}}
\end{equation}
for some $C>0$.

\noindent {\bf \ref{AS}.3} For $y\geq \e N^{2/3}$, $\e$ as in (\ref{Asympt2}),
\begin{equation}\label{Asympt3}
|Z(a) \kappa N^{1/3} I_{m,d}(u+y \kappa N^{1/3})| \leq C e^{-\frac12 \e^{1/2}  y N^{1/3}}
\end{equation}
for some $C>0$.

\subsection{Asymptotics of $\tilde I_{m,d}(u+y \kappa N^{1/3})$, $0<\rho<1$}\label{AS2}
{\bf \ref{AS2}.1}  For fixed $y\in \R$,
\begin{equation}\label{Asympt4}
Z(a) \kappa N^{1/3} \tilde I_{m,d}(u+y \kappa N^{1/3}) = 2\pi i \Ai(w^2+s+y) e^{-w(w^2+s+y)} e^{\frac13 w^3}+\Or(N^{-1/3}).
\end{equation}

\noindent {\bf \ref{AS2}.2} For $0\ll L \leq y \leq \e N^{2/3}$, $\e>0$ small enough, $L>0$ large enough,
\begin{equation}\label{Asympt5}
|Z(a) \kappa N^{1/3} I_{m,d}(u+y \kappa N^{1/3})| \leq  C e^{-\frac13  y^{3/2}}
\end{equation}
for some $C>0$.

\noindent {\bf \ref{AS2}.3} For $y\geq \e N^{2/3}$, $\e$ as in (\ref{Asympt5}),
\begin{equation}\label{Asympt6}
|Z(a) \kappa N^{1/3} I_{m,d}(u+y \kappa N^{1/3})| \leq C e^{-\frac12 \e^{1/2}  y N^{1/3}}
\end{equation}
for some $C>0$.

\begin{proofOF}{(\ref{Asympt1})}
We have to estimate
\begin{eqnarray}
I_{N+\alpha/2,\alpha/2}(u+y \kappa N^{1/3})&=&\int_{\Gamma_{1-\rho}} \dx z e^{-z (u+y \kappa N^{1/3})} \frac{(z+\rho)^{N}}{(1-\rho-z)^{N+\alpha}}\\
&=& \int_{\Gamma_{1-\rho}} \dx z e^{N f_N(z)} \nonumber
\end{eqnarray}
with
\begin{equation}
 f_N(z)=-z(u/N+y \kappa N^{-2/3})+\ln(z+\rho)-(1+\alpha/N)\ln(1-\rho-z)
\end{equation}
for any fixed $y\in \R$.

Let us define
\begin{equation}
f_\infty(z)=\lim_{N\to\infty} f_N(z)=-\frac{z}{\rho^2}+\ln(z+\rho)-\frac{(1-\rho)^2}{\rho^2}\ln(1-\rho-z)
\end{equation}
and find a steep descent path for it which is close to the steepest descent one for $z$ close to the critical point, which is the solution of
\begin{equation}
\Dt{f_\infty(z)}{z}=-\frac{1}{\rho^2}+\frac{1}{z+\rho}+\frac{(1-\rho)^2}{\rho^2}\frac{1}{1-\rho-z}=0.
\end{equation}
There is a double solution for $z=0$, thus $f_\infty''(z=0)=0$. Moreover,
\begin{equation}
\Dttt{f_\infty(z)}{z}\Big|_{z=0}=\frac{2}{\rho^3(1-\rho)}.
\end{equation}
Therefore we choose as integration path the one shown in Figure~\ref{figFixed}.
\begin{figure}[t!]
\begin{center}
\psfrag{g1}{$\gamma_1$}
\psfrag{g2}{$\gamma_2$}
\psfrag{g3}{$\gamma_3$}
\psfrag{1-r}[c]{$1-\rho$}
\psfrag{p/3}{$\pi/3$}
\includegraphics[height=5cm]{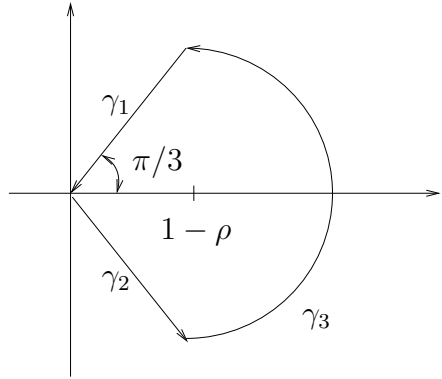}
\caption{Integration path used for the asymptotics for fixed $y$.}\label{figFixed}
\end{center}
\end{figure}
The chosen path is a steep descent path for $f_\infty$ as is discussed now.

The path $\gamma_2$ is given by $\{z=te^{-i \pi/3}$,$t\in[0,2(1-\rho)]\}$. The real part of $f_\infty$ on $\gamma_2$ is then
\begin{equation}
Re(f_\infty)= -\frac{t}{2\rho^2} +\frac12 \ln\Big((\rho+\tfrac12 t)^2+\frac34 t^2\Big)-\frac{(1-\rho)^2}{2\rho^2} \ln\Big((1-\rho-\tfrac12 t)^2+\frac34 t^2\Big).
\end{equation}
Therefore
\begin{equation}
\Dt{Re(f_\infty)}{t} = -\frac{t^2 (2\rho(1-\rho)+t(1-2\rho)+t^2)}{2\rho^2(\rho^2+\rho t+ t^2)((1-\rho-t)^2+t(1-\rho))}.
\end{equation}
The denominator is positive and it is easy to see that the numerator is always strictly positive for $t\in(0,2(1-\rho)]$ and for all $\rho\in (0,1)$. Therefore $\gamma_2$ is a steep descent path, and by symmetry $\gamma_1$ is a steep ascent path.

\noindent The path $\gamma_3$ is given by $\{z=1-\rho+\sqrt{3}(1-\rho) e^{i\varphi}, \varphi\in [-\pi/2,\pi/2]\}$. On $\gamma_3$
\begin{eqnarray}
Re(f_\infty)&=& -\frac{(1-\rho)\sqrt{3}\cos\varphi}{\rho^2} +\frac12 \ln(1+3(1-\rho)^2+2\sqrt{3}(1-\rho)\cos\varphi) \nonumber\\
& &-\frac{(1-\rho)^2}{\rho^2} \ln(\sqrt{3}(1-\rho)),
\end{eqnarray}
which implies
\begin{eqnarray*}
\Dt{Re(f_\infty)}{\cos\varphi} &=& -\frac{(1-\rho)\sqrt{3}}{\rho^2} + \frac{\sqrt{3}(1-\rho)}{1+3(1-\rho)^2+2\sqrt{3}(1-\rho)\cos\varphi} \\
&=& -\frac{(1-\rho)\sqrt{3}(2\sqrt{3}(1-\rho)\cos\varphi + 3 (1-\rho)^2+1-\rho^2)}{\rho^2 (1+3(1-\rho)^2+2\sqrt{3}(1-\rho)\cos\varphi)}<0.
\end{eqnarray*}
Thus the path of Figure~\ref{figFixed} is a steep descent path $\Gamma_{1-\rho}$.

The first consequence is the following. Denote by $\Gamma_{1-\rho}(\delta)=\Gamma_{1-\rho}\big|_{|z|\leq \delta}$ the part of the path $\Gamma_{1-\rho}$ closer than $\delta$ to the origin. Then for any $\delta>0$ and $N$ large enough,
\begin{equation}
\bigg|\int_{\Gamma_{1-\rho}} e^{N f_N(z)}\dx z - \int_{\Gamma_{1-\rho}(\delta)}e^{N f_N(z)}\dx z \bigg| \leq e^{N f_N(0)} \Or(e^{-\mu N})
\end{equation}
for some $\mu=\mu(\delta)\sim \delta^3>0$. Remark that
\begin{equation}
e^{N f_N(0)}=Z(a)^{-1}.
\end{equation}
Consequently we need to estimate the integral close to $z=0$ on $\Gamma_{1-\rho}(\delta)$ only. We use the Taylor expansion,
\begin{eqnarray}
f_N(z) &=& f_N(0)+f_N'(0) z +\frac12 f_N''(0) z^2+ \frac16 f_N'''(0) z^3\\
&& + \Or\big(|z|^4 \max_{0\leq |z| \leq \delta}|f_N^{(iv)}(z)|\big).
\end{eqnarray}
Some computations yield
\begin{eqnarray}
f_N'(0)&=&-N^{-2/3} (y+s)\kappa+\Or(N^{-1}),\nonumber \\
f_N''(0)&=&N^{-1/3} 2 w \kappa^2+\Or(N^{-2/3}), \\
f_N'''(0)&=&2\kappa^3+\Or(N^{-1/3}), \nonumber
\end{eqnarray}
and $|f_N^{(iv)}(z)|=\Or(1)$ for $|z|\leq \delta$. The change of variable $\tau=N^{1/3} \kappa z$ leads to
\begin{equation}
N f_N(z)=N f_N(0)-\tau (y+s) +w \tau^2+\frac13 \tau^3 + \Or(\tau,\tau^4) N^{-1/3}.
\end{equation}
Consequently
\begin{eqnarray}\label{eq6}
&&e^{-N f_N(0)}\int_{\Gamma_{1-\rho}(\delta)} e^{N f_N(z)}\dx z = \kappa^{-1} N^{-1/3}\int_{\kappa N^{1/3} \Gamma_{1-\rho}(\delta)}\dx \tau e^{-\tau (y+s) +w\tau^2+\frac13 \tau^3}\nonumber \\
&+&  \kappa^{-1} N^{-1/3}\int_{\kappa N^{1/3} \Gamma_{1-\rho}(\delta)} \dx \tau e^{-\tau (y+s) +w \tau^2+\frac13 \tau^3} \Big(e^{\Or(\tau,\tau^4)N^{-1/3}}-1\Big).
\end{eqnarray}
The last term can be estimated using that $|e^x-1|\leq |x| e^{|x|}$, i.e., using
\begin{equation}
\Big(e^{\Or(\tau,\tau^4)N^{-1/3}}-1\Big) = e^{\Or(\tau,\tau^4)N^{-1/3}} \Or(\tau,\tau^4)N^{-1/3} .
\end{equation}
The term in the exponent is of the form $-\tau (y+s)\chi_1 +w \tau^2\chi_2+\frac13 \tau^3\chi_3$ for some $\chi_1,\chi_2,\chi_3$. By taking $\delta$ small enough $\chi_1,\chi_2,\chi_3$ can be made  as close to $1$ as desired. Thus the second integral converges and the error term in (\ref{eq6}) is of order $\Or(N^{-2/3})$.

Finally we estimate the leading term
\begin{equation}
\kappa^{-1} N^{-1/3} \int_{\kappa N^{1/3} \Gamma_{1-\rho}(\delta)} \dx \tau e^{-\tau (y+s) +w \tau^2+\frac13 \tau^3}.
\end{equation}
Deforming the integration path from $\Gamma(\delta) N^{1/3}\kappa$ to $\Gamma(\delta) N^{1/3}\kappa-w$, one obtains
\begin{equation}
e^{(y+s)w}e^{\frac23 w^3}\kappa^{-1} N^{-1/3} \int_{\Gamma(\delta) N^{1/3}\kappa} \dx \tau e^{-\tau (y+s+w^2)+\frac13 \tau^3}
\end{equation}
up to an $\Or(e^{-\mu N})$ error. By extending the integral to $e^{\pm i\pi/3}\infty$ one picks up an error at most $\Or(e^{-\mu N})$, again. But
\begin{equation}
\int_{\Gamma(\infty)} \dx z e^{\frac13 z^3-x z} = -2\pi i \Ai(x),
\end{equation}
where $\Gamma(\infty)$ is the path joining $0$ with $e^{\pm i\pi/3}\infty$ by straight lines oriented with imaginary part decreasing. Note that in (\ref{Airy2}) the orientation is the opposite. The error term $\Or(e^{-\mu N})$ can be bounded by $\Or(N^{-1/3})$. Thus putting all the terms together we have proved that for any \emph{fixed} $y$
\begin{eqnarray}
I_{N+\alpha/2,\alpha/2}(u+y \kappa N^{1/3}) Z(a) \kappa N^{1/3}& =&-2\pi i \Ai(w^2+s+y)e^{w(w^2+s+y)}e^{-\frac13 w^3}\nonumber \\
& &+\Or(N^{-1/3}).
\end{eqnarray}
\end{proofOF}

\begin{proofOF}{(\ref{Asympt2})}
Let us define $\tilde y=y \kappa N^{-2/3} \in [L \kappa N^{-2/3},\kappa \e]$,
\begin{equation}
f_N(z)=-z (u/N+\tilde y)+\ln(\rho+z)-(1+\alpha/N)\ln(1-\rho-z),
\end{equation}
and
\begin{equation}
g(z)=-z (\rho^{-2}+\tilde y)+\ln(\rho+z)-\frac{(1-\rho)^2}{\rho^2}\ln(1-\rho-z).
\end{equation}
$g$ has a real positive critical point at
\begin{equation}
z_c=\tilde y^{1/2} \kappa^{-3/2}+\Or(\tilde y).
\end{equation}
Let $z_0=\tilde y^{1/2} \kappa^{-3/2}$. Then as integration path we choose $\Gamma_{1-\rho}=\gamma_1\wedge \gamma_2$, where $\gamma_1=\{z | z=z_0-i t, t\in[\sqrt3 z_0,\sqrt3 z_0]\}$, $\gamma_2$ the path used in the case of fixed $y$ restricted to $Re(z)>z_0$, see Figure~\ref{figLarge}.
\begin{figure}[t!]
\begin{center}
\psfrag{g1}[c]{$\gamma_1$}
\psfrag{g2}{$\gamma_2$}
\psfrag{1-r}[c]{$1-\rho$}
\psfrag{z0}{$z_0$}
\includegraphics[height=5cm]{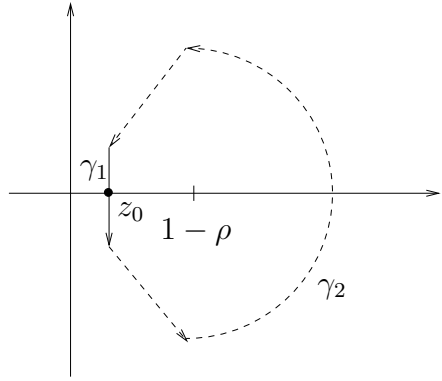}
\caption{Integration path used for the asymptotics for large value of $y$. $\gamma_2$ is the dashed line.}\label{figLarge}
\end{center}
\end{figure}
Since $\tilde y>0$, the path $\gamma_2$ is steep descent because it is for $\tilde y=0$ and $Re(z)>0$ on $\gamma_2$. Thus we only have to check it on $\gamma_1$. On $\gamma_1$ we have
\begin{equation}
Re(f_N(z))= -z_0\Big(\frac{u}{N}+\tilde y\Big)+\frac12 \ln\big((\rho+z_0)^2+t^2\big)-\frac12\Big(1+\frac{\alpha}{N}\Big)\ln\big((1-\rho-z_0)^2+t^2\big)
\end{equation}
and
\begin{equation}
\Dt{Re(f_N)}{t} = -t\frac{(t^2+(\rho+z_0)^2)\alpha/N+2\rho-1+2z_0}{\big((\rho+z_0)^2+t^2\big)\big((1-\rho-z_0)^2+t^2\big)}.
\end{equation}
The denominator is obviously positive.

Next consider the numerator
\begin{equation}
M=(t^2+(\rho+z_0)^2)\alpha/N+2\rho-1+2z_0.
\end{equation}
For $\rho\in (0,1/2)$, $\alpha/N= (1-2\rho)/\rho^2+\Or(N^{-1/3})$, $M$ can be rewritten as $$M=\alpha/N t^2+(1-2\rho)((1+z_0/\rho)^2-1)+2 z_0+\Or(N^{-1/3})>0$$ for $N$ large enough. For $\rho \in (1/2,1)$, $\alpha= (1-2\rho)\rho^2 + \Or(w N^{-1/3})$ and $M$ can be rewritten as
$$M=\frac{\alpha}{N} t^2 + 2 z_0 \frac{1-\rho}{\rho}+z_0^2\frac{1-2\rho}{\rho^2} +\Or(w N^{-1/3}).$$ Since $z_0 = \tilde y^{1/2} \kappa^{-3/2} \geq L^{1/2} \kappa^{-1} N^{-1/3}$, and that $z_0 \leq \e \kappa^{-1}\ll 1$, both the quadratic term in $z_0$ and the $\Or(w N^{-1/3})$ are dominated by $2 z_0 (1-\rho)/\rho$ for $L\gg 1$, $\e\ll 1$, and $N$ large enough. Thus $M(t=0)\geq L^{1/2} \kappa^{-1} N^{-1/3}>0$ for $L$ large enough. By monotonicity of $M$ in $t$ we have to check that $M(t=\sqrt3 z_0)>0$, the maximal value which $t$ takes in $\gamma_1$. But $M(t=\sqrt3 z_0)- M(t=0) \sim z_0^2$, which is dominated by the linear term in $z_0$ again. Thus $M(t)>0$ for $N$ large enough and for all $t\in [-\sqrt3 z_0,\sqrt3 z_0]$.

We have shown that for all $\rho\in (0,1)$, $\Dt{Re(f_N)}{t}<0$ for $t>0$, $\Dt{Re(f_N)}{t}>0$ for $t<0$ for $L\gg 1$, $\e \ll 1$, and $N$ large enough. Thus $\gamma_1$ is a steep descent path. Therefore, if we denote by $\Gamma_{1-\rho}(\delta)^c$ the portion of $\Gamma_{1-\rho}$ with $|z-z_0|>\delta$,
\begin{equation}\label{eq7}
\bigg|\int_{\Gamma_{1-\rho}(\delta)^c} \dx z e^{N f_N(z)}\bigg| \leq e^{N f_N(z_0)} \Or(e^{-\mu N})
\end{equation}
for some $\mu>0$.

Finally we have to evaluate the contribution coming from $\Gamma_{1-\rho}(\delta)$, the portion of $\Gamma_{1-\rho}$ with $|z-z_0|\leq \delta$. The contribution of the part in $\gamma_1$ is estimated as follows. On $\gamma_1$,
\begin{equation}
Re(f_N(z))= f_N(z_0) -\frac{t^2}{2} f_N''(z_0) + \Or(t^4).
\end{equation}
Some computations leads to $f_N''(z_0) = \Or(w N^{-1/3}) + (2\kappa^{3/2} + \Or(w N^{-1/3}))\tilde y^{1/2}+\Or(\tilde y).$ For $L\gg 1$, $\e\ll 1$, and $N$ large enough, it follows that $f_N''(z_0) \geq \frac32 \kappa^{3/2} \tilde y^{1/2}$. On the other hand, the term proportional to $t^4$ is much smaller than the $t^2$ term. In fact, for $0\leq t \leq \sqrt3 z_0\sim \tilde y^{1/2}$, $t^2 \leq \tilde y\Or(1) \leq \tilde y^{1/2} \Or(\e^{1/2}) \ll f_N''(z_0)$. Therefore
\begin{eqnarray}\label{eq8}
\bigg|\int_{\gamma_1} \dx z e^{N f_N(z)}\bigg| &\leq& e^{N f_N(z_0)} \int_\R \dx t e^{-\frac13 t^2 f_N''(z_0) N} \leq e^{N f_N(z_0)} \int_\R \dx t e^{-\frac12 t^2 \kappa^2 y^{1/2} N^{2/3}} \nonumber \\
&=& e^{N f_N(z_0)}  y^{-1/4} N^{-1/3} \Or(1).
\end{eqnarray}

In (\ref{eq7}) and (\ref{eq8}) we still have to evaluate $f_N(z_0)-f_N(0)$. A computation leads to
\begin{equation}
N f_N(z_0)-N f_N(0) = - s y^{1/2} \chi_1 + w y \chi_2 - \frac23 y^{3/2} \chi_3
\end{equation}
for some $\chi_1,\chi_2,\chi_3$ which can be made as close to $1$ as wanted by choosing $\e$ small enough. On the other hand, since we have $y\geq L$, for $L$ large enough,
\begin{equation}
N f_N(z_0)-N f_N(0) \leq -\frac13 y^{3/2}.
\end{equation}
Thus the contributions (\ref{eq7}) and (\ref{eq8}) can be bounded by
\begin{eqnarray}\label{eq10}
(\ref{eq7}) &\leq & Z(a)^{-1} e^{-\mu N} e^{-\frac13 y^{3/2}} \Or(1), \nonumber \\
(\ref{eq8}) &\leq & Z(a)^{-1} N^{-1/3} e^{-\frac13 y^{3/2}} L^{-1/4} \Or(1).
\end{eqnarray}

The final step is to bound the contribution coming form $\gamma_2\cup \Gamma_{1-\rho}(\delta)$. From the asymptotics of fixed $y$ one has, using $z=t e^{i\pi/3}$,
\begin{eqnarray}
Re(f_N(z)) &=& f_N(0)-(\tilde y + s \kappa N^{-2/3}) \frac{t}{2} + \Or(t N^{-1})  \\
& &+ w \kappa^2 N^{-1/3} \frac{t^2}{2}+\Or(t^2 N^{-2/3})-\frac{\kappa}{3} t^3 +\Or(t^3 N^{-1/3},t^4).\nonumber
\end{eqnarray}
In this case, the parameter $t$ takes values in $0<2 z_0 \leq t \leq \delta/\sqrt{3} \ll 1$. Moreover recall that $z_0 \geq L^{1/2} \kappa^{-1} N^{-1/3}$. In the term linear in $t$, $\tilde y$ dominates the others for large $L$. For the minimal value taken by $t$, the quadratic term is $\sim L N^{-1}$, and the cubic term is $\sim L^{3/2} N^{-1}$. Thus for large $L$, the cubic term dominates the quadratic one. But since $t\leq \delta/\sqrt3$, the quartic term is also dominated by the cubic one. Therefore,
\begin{equation}
Re(f_N(z)) \leq f_N(0) -\frac{ \tilde y t}{4}- \frac{\kappa t^3}{6} \leq f_N(0)-\frac{\tilde y^{3/2}}{2\kappa^{3/2}}-\frac{\kappa t^3}{6}.
\end{equation}
Thus
\begin{equation}\label{eq9}
\bigg|\int_{\gamma_2\cup\Gamma_{1-\rho}(\delta)} \hspace{-0.5cm}\dx z e^{N f_N(z)}\bigg| \leq 2 e^{N f_N(0)} e^{-\frac12 y^{3/2}} \int_0^\infty \dx t e^{-\frac16 \kappa t^3 N} \leq \frac{Z(a)^{-1} e^{-\frac12 y^{3/2}}}{N^{1/3}} \Or(1).
\end{equation}
From (\ref{eq10}) and (\ref{eq9}) the desired bound follows.
\end{proofOF}

\begin{proofOF}{(\ref{Asympt3})}
The proof of this bound is based on the estimate (\ref{Asympt2}). We use (\ref{Asympt2}) for $\tilde y=\e\kappa/2$ with the result
\begin{equation}
f_N(z)=f_N(z)\big|_{\tilde y=\e\kappa/2} - z (\tilde y-\e\kappa/2) \leq f_N(z)\big|_{\tilde y=\e\kappa/2} - 2^{-3/2} \sqrt{\e} y N^{-2/3},
\end{equation}
because $z\geq \sqrt{\e/2}\kappa^{-1}$ and $\tilde y-\e\kappa/2 \geq \tilde y/2$. It follows that
\begin{equation}
\Big|I_{N+\alpha/2,\alpha/2}(u+y\kappa N^{1/3}) \kappa N^{1/3} Z(a)\Big| \leq C e^{-\e^{3/2} N/2}e^{-\sqrt{\e} y N^{1/3}/4} \leq C e^{-\sqrt{\e} y N^{1/3}/4}.
\end{equation}
\end{proofOF}

\begin{proofOF}{(\ref{Asympt4}), (\ref{Asympt5}), and (\ref{Asympt6})}
The asymptotics of $\tilde I$ are similar to the one of $I$. Instead of computing everything again, we show that, via the transformation $\rho\mapsto 1-\rho$ and $w\mapsto -w$, one obtains essentially the same integrals as already studied for $I$. More precisely, we have to estimate the asymptotics of
\begin{eqnarray}
W(N,\rho,w)&=&Z(a) \kappa N^{1/3}I_{N+\alpha/2,\alpha/2}(u+y \kappa N^{1/3}) \\
&=& \int_{\Gamma_{1-\rho}} \dx z e^{-z(u(N,\rho,w)+y\kappa(\rho) N^{1/3})}\kappa(\rho)N^{1/3} \frac{\Big(1+\frac{z}{\rho}\Big)^N}{\Big(1-\frac{z}{1-\rho}\Big)^{N+\alpha(N,\rho,w)}}\nonumber
\end{eqnarray}
and
\begin{eqnarray}
\widetilde W(N,\rho,w)&=&Z(a)^{-1} \kappa N^{1/3} \tilde I_{N+\alpha/2,\alpha/2}(u+y \kappa N^{1/3}) \\
&=& \int_{\Gamma_{-\rho}} \dx z e^{z(u(N,\rho,w)+y\kappa(\rho) N^{1/3})}\kappa(\rho)N^{1/3} \frac{\Big(1-\frac{z}{1-\rho}\Big)^{N+\alpha(N,\rho,w)}}{\Big(1+\frac{z}{\rho}\Big)^{N}}.\nonumber
\end{eqnarray}
Here the dependence in $N,\rho,w$ of $u,\alpha,\kappa$  is displayed explicitly, since it is needed below. A simple calculation shows that
\begin{eqnarray}
& &\widetilde W(N,1-\rho,-w) \\
&=&-\int_{\Gamma_{1-\rho}} \dx z e^{-z(u(N,1-\rho,-w)+y\kappa(1-\rho) N^{1/3})}\kappa(1-\rho)N^{1/3} \frac{\Big(1-\frac{z}{\rho}\Big)^{N+\alpha(N,1-\rho,-w)}}{\Big(1+\frac{z}{1-\rho}\Big)^{N}}.\nonumber
\end{eqnarray}
Let us define $M=N+\alpha(N,1-\rho,-w)$, then with an explicit but lengthy computations one establishes
\begin{eqnarray}\label{eq11}
u(N,1-\rho,-w)&=&u(M,\rho,w)+\Or(1),\nonumber \\
\kappa(1-\rho) N^{1/3} &=& \kappa(\rho) M^{1/3} + \Or(1), \\
N&=& M + \alpha(M,\rho,w)+\Or(1).\nonumber
\end{eqnarray}
(\ref{eq11}) implies that in the asymptotics of $\widetilde W(N,1-\rho,-w)$ are the same as the asymptotics of $-W(M,\rho,w)$, since the corrections of $\Or(1)$ in (\ref{eq11}) are negligible for the asymptotics (\ref{Asympt1}), (\ref{Asympt2}), and (\ref{Asympt3}).
\end{proofOF}

\appendix

\section{Scaling functions}\label{AppBRlink}
The scaling function $g(s,w)$ defined in (\ref{eqScalingG}) is precisely the one derived in (\ref{eqFinal}) except for the shift to $s\mapsto s-w^2$, since in (\ref{eqFinal}) $g(s+w^2,w)$ is obtained. Below we establish that $g$ is identical to the Baik-Rains scaling function $g_{\rm BR}$~\cite{BR00}. $g(s,w)$ is continuous at $w=0$ and even in $w$. $g_{\rm BR}(s,w)$ has the same properties. Thus it suffices to consider $w>0$. We first rewrite (\ref{eqFinal}) in another form by moving the factor $s$ from the integrand to the limit of the integral, since later on we have to deal with the derivatives $\frac{\partial}{\partial s}g(s,w)$. Define the functions
\begin{eqnarray}
\widetilde \Phi_s(x) &=&\int_0^\infty \dx y \Ai(x+y)\int_{-\infty}^s \dx z e^{w z} \Ai(y+z),\nonumber\\
\widetilde \Psi_w(x) &=&\int_{-\infty}^0\dx y e^{w y}\Ai(x+y),\\
\widetilde \rho_s(x,y)&=&(\Id-P_s K_{\Ai}P_s)^{-1}(x,y).\nonumber
\end{eqnarray}
Then by using the representation (\ref{eqSbis}) for $S_{w,s}$, (\ref{eqFinal}) rewrites as
\begin{eqnarray}\label{A.2}
 g(s,w)&=& e^{-\frac13 w^3}\bigg[ \int_{-\infty}^s\dx x \int_{-\infty}^0 \dx y \Ai(x+y)e^{w(x+y)}\nonumber \\
& &\hspace{26pt}+\int_s^\infty\dx x \int_s^\infty \dx y \widetilde \Phi_s(x) \widetilde \rho_s(x,y)\widetilde \Psi_w(y)\bigg].
\end{eqnarray}
The relations with the functions of (\ref{eqFinal}) are $\widetilde \Phi_s(x)=\Phi_{w,s-w^2}(x-s)$, $\widetilde \Psi_w(x)=e^{\frac13 w^3}\Psi_{w,s-w^2}(x-s)$, and $\widetilde \rho_s(x,y)=(\Id-P_0 K_{\Ai,s} P_0)^{-1}(x-s,y-s).$

Baik and Rains~\cite{BR00}, see also~\cite{Pra03}, use the Riemann-Hilbert techniques and arrive at a limit function, $g_{\rm BR}$, which is given as the solution of a set of differential equations\footnote{Comparing with the functions in Lemma 3.1 of~\cite{BR00}, one sees that there has been a change from $2w$ to $w$, so that for example, the function $a(s,w)$ below equals to the function $a(s,w/2)$ in~\cite{BR00}.}. More precisely, with $a=a(s,w)$, $b=b(s,w)$, $q=q(s)$, one considers
\begin{eqnarray}\label{A.4}
\frac{\partial}{\partial s}a &=& q b,\nonumber\\
\frac{\partial}{\partial s}b &=& q a -w b,
\end{eqnarray}
and
\begin{eqnarray}\label{A.5}
\frac{\partial}{\partial w}a &=& q^2 a-(q'+wq) b,\nonumber\\
\frac{\partial}{\partial w}b &=& (q'-wq) a+(w^2-s-q^2) b.
\end{eqnarray}
Here $q=q(s)$ is the Hastings-McLeod solution to the Painlev\'{e}
II equation
\begin{equation}\label{A.6}
q'' = 2q^3 + sq,
\end{equation}
which is singled out by the condition $q(s)<0$ for all $s\in\R$.
The Hastings-McLeod solution has the asymptotics $q(s)\cong-\Ai(s)$ for $s\to\infty$ and $u(s)\cong-(-s/2)^{1/2}$
for $s\to-\infty$. (\ref{A.4}) and (\ref{A.5}) have to be solved
with the initial condition
\begin{equation}\label{A.7}
a(s,0)=-b(s,0)=\exp \Big(\int^\infty_s \dx s' q(s')\Big).
\end{equation}
The Baik-Rains scaling function is defined through
\begin{equation}\label{A.8}
g_{\mathrm{BR}}(s,w)= \int^s_{-\infty} \dx s' a(s',w) a(s',-w).
\end{equation}
\begin{proposition}\label{A.prop1}
With the above definitions
\begin{equation}\label{A.9}
g(s,w)= g_{\mathrm{BR}}(s,w).
\end{equation}
\end{proposition}
\begin{proof} We fix $w>0$. We will establish that
\begin{equation}\label{A.9a}
 \frac{\partial}{\partial s}g(s,w)=a(s,w) a(s,-w).
\end{equation}
Then
\begin{equation}\label{A.10}
 g(s,w)= g_{\mathrm{BR}}(s,w)+ c.
\end{equation}
Now, by construction,
\begin{equation}\label{A.11}
F_w(s)\quad\mathrm{and}\quad F^{\mathrm{BR}}_w (s)= \frac{\partial}{\partial s}\big(F_{\rm GUE}(s+w^2)g_{\mathrm{BR}}(s+w^2,w)\big)
\end{equation}
are distribution functions with mean zero. From (\ref{A.10}) we infer
\begin{equation}\label{A.12}
\int_\R s \dx F_w(s)= \int_\R s \dx F^{\mathrm{BR}}_w(s)+c \int_\R \dx s \, s
\Dtt{}{s} F_{\rm GUE}(w^2+s).
\end{equation}
Since $F_{\rm GUE}(s)$ is also a distribution function and $\Dt{}{s}F_{\rm GUE}(s)$ has a fast decay at infinity, (\ref{A.12}) amounts to $0=0-c$ and thus $c=0$.

To check (\ref{A.9a}) we will differentiate $\tilde{g}=e^{w^3/3}g$. It is convenient to follow the scheme in \cite{SI04}, Section 4.2, according to which
\begin{equation}\label{A.13}
a(s,\pm w)=1-\int^\infty_s \dx x \int^\infty_s \dx y \Ai(x) \widetilde \rho_s(x,y) \widetilde \Psi_{\pm w}(y).
\end{equation}
Remark that for $w>0$ one can rewrite $\widetilde \Psi_w(x)=\int_{-\infty}^0\dx y e^{w y}\Ai(x+y)$  using (\ref{Airy3}).

We differentiate as
\begin{eqnarray}\label{A.17}
\frac{\partial}{\partial s}\tilde{g}(s,w) &=& e^{w s} \widetilde \Psi_w(s)
-\widetilde \Phi_s(s) \int^\infty_s \dx y \widetilde \rho_s(s,y)\widetilde \Psi_w(y) \nonumber \\
& &-\widetilde \Psi_w(s)\int^\infty_s \dx x \widetilde \Phi_s(x) \widetilde \rho_s(x,s)\nonumber\\
 & & +\int^\infty_s \dx x \int^\infty_s \dx y \frac{\partial \widetilde \Phi_s(x)}{\partial s}\widetilde \rho_s(x,y)\widetilde \Psi_w(y)\nonumber\\
 & &+ \int^\infty_s \dx x \int^\infty_s \dx y
 \widetilde \Phi_s(x)\frac{\partial \widetilde \rho_s(x,y)}{\partial s} \widetilde \Psi_w(y).
\end{eqnarray}
The central identity for the proof is, see~\cite{SI04,TW94},
\begin{equation}\label{A.19}
\frac{\partial}{\partial s}\widetilde \rho_s(x,y)=-\frac{\partial}{\partial x}\widetilde \rho_s(x,y) -\frac{\partial}{\partial y}\widetilde\rho_s(x,y)-Q(x)Q(y)
\end{equation}
with
\begin{equation}\label{A.20}
Q(x)=\int^\infty_s \dx y \widetilde\rho_s(x,y)\Ai(y).
\end{equation}
We insert (\ref{A.19}) in (\ref{A.17}) and integrate by parts to eliminate the terms $\frac{\partial}{\partial x}\widetilde\rho_s(x,y)$ and $\frac{\partial}{\partial y}\widetilde\rho_s(x,y)$, which requires the derivatives
\begin{eqnarray}
\frac{\partial}{\partial x}\widetilde \Psi_w(x) &=& \Ai(x)-w \widetilde \Psi_w(x),\\
\frac{\partial}{\partial x}\widetilde \Phi_s(x) &=& w\widetilde \Phi_s(x)-\Ai(x)
\int^s_{-\infty} \dx y \Ai(y)e^{wy}-\frac{\partial}{\partial s}\widetilde \Phi_s(x).\nonumber
\end{eqnarray}
In the end only four terms remain, which can be assembled as
\begin{equation}\label{A.21}
\frac{\partial}{\partial s}\tilde{g}(s,w) =\Big(1-\int^\infty_s \dx x\widetilde \Psi_w(x) Q(x)\Big) \Big(\int^s_{-\infty} \dx x\Ai(x)e^{wx}+\int^\infty_s
\dx x \widetilde \Phi_s(x)Q(x)\Big).
\end{equation}

The first factor in (\ref{A.21}) equals $a(s,w)$. To prove Proposition~\ref{A.prop1} it thus remains to establish that the second factor equals $e^{\frac13 w^3} a(s,-w)$. Let $\psi_{-w}(x)=e^{wx}$. Then
\begin{equation}\label{A.22}
\widetilde \Phi_s = K_{\Ai} \psi_{-w} +(\Id-K_{\Ai})P_s \psi_{-w} - P_s \psi_{-w}
\end{equation}
and
\begin{equation}\label{A.23}
Q(x)=(\Id-P_s K_{\Ai}P_s)^{-1}P_s \Ai(x).
\end{equation}
Thus the second factor in (\ref{A.21}) writes
\begin{eqnarray}\label{A.24}
\langle \psi_{-w},Q_s \Ai\rangle + \langle \widetilde \Phi_s , P_s Q\rangle
&=& \langle \psi_{-w}, Q_s \Ai\rangle + \langle K_{\Ai}\psi_{-w} , P_s Q\rangle -\langle \psi_{-w},P_s Q\rangle \nonumber \\
& & + \langle \psi_{-w}, P_s (\Id-K_{\Ai})P_s (\Id-P_s K_{\Ai} P_s)^{-1} P_s \Ai\rangle \nonumber \\
&=& \langle \psi_{-w},\Ai\rangle +\langle K_{\Ai}\psi_{-w} , P_s Q\rangle\nonumber \\
& & -\langle \psi_{-w},P_s Q\rangle,
\end{eqnarray}
since the last term in the middle part equals $\langle \psi_{-w},P_s \Ai\rangle$. According to (\ref{B.13})
\begin{equation}\label{A.26}
e^{\frac13 w^3} a(s,-w)= e^{\frac13 w^3}-\langle e^{\frac13 w^3}\widetilde \Psi_{-w},P_s Q\rangle
\end{equation}
and by (\ref{Airy3}) it follows that $e^{\frac13 w^3}=\langle\psi_{-w},\Ai\rangle$. Moreover,
\begin{equation}
e^{\frac13 w^3}\widetilde \Psi_{-w}(x)=\psi_{-w}(x)-e^{\frac13 w^3} \int_0^\infty\dx y e^{-w y}\Ai(x+y).
\end{equation}
An explicit computation shows that $e^{\frac13 w^3} \int_0^\infty\dx y e^{-w y}\Ai(x+y)=(K_{\Ai} \psi_{-w})(x)$. Thus
\begin{equation}\label{A.25}
e^{\frac13 w^3}\widetilde \Psi_{-w}=\psi_{-w}-K_{\Ai}\psi_{-w}.
\end{equation}
Inserting (\ref{A.25}) in (\ref{A.26}) one establishes that (\ref{A.24}) equals $e^{\frac13 w^3}a(s,-w)$.
\end{proof}

\section{Determinantal fields: proof of Propositions \ref{PropInverse} and \ref{Prop4}}\label{AppProp4}

\subsection{No boundary sources}\label{sectB1}
To prove Proposition~\ref{Prop4} we find it computationally convenient to approximate the exponential distribution through a geometric one. Then $w(i,j)$, $i,j\geq 1$, are independent random variables with $\Pb(\{w(i,j)=n\})=(1-q)q^n$, $0<q<1$. The RSK construction, as explained in the main text, can be carried through with minor modifications, compare with~\cite{SI04}. The lines $j\mapsto h_\ell(j,\tau)$ take values in $\Z$ and are pinned as $h_\ell(\pm\tau,\tau)=\ell$, $\ell=0,-1,\ldots$. The weight of a jump of size $\delta$ is $(\sqrt{q})^{|\delta|}$. Let us denote the corresponding point random field by $\phi^q_\tau(j,n)$, $|j|\leq \tau$, $n\in\Z$, i.e.,
\begin{equation}\label{B.1}
\phi^q_\tau(j,n)=
 \begin{cases}
 1 & \textrm{if there is an }\ell\textrm{ such that }h_\ell(j,\tau)=n,\\
 0 & \textrm{otherwise}.
 \end{cases}
\end{equation}
It is determinantal and, at equal times $j=2d$, $\tau=2m$,
\begin{equation}\label{B.2}
\E\Big(\prod^{M}_{k=1}\phi^q_\tau(2d,n_k)\Big)=
\det\big(K^q_{m,d}(n_k,n_{k'})\big)_{1\leq k,k'\leq M}.
\end{equation}
We will first compute the kernel $K^q_{m,d}$ and then show that in
the exponential limit, lattice spacing $\e$,
$q=1-\e$,
\begin{equation}\label{B.3}
\lim_{\e\to
0}\e^{-1}K^{1-\e}_{m,d}\big(\lfloor \e^{-1}x\rfloor,
\lfloor\e^{-1}x'\rfloor\big)= K_{m,d}(x,x')
\end{equation}
for $x,x'>0$.

We have to compute $K^q_{m,d}$, for which we use the Fermion formalism of~\cite{PS02}. Let $[-N,-N+1,\ldots,N]=\Lambda_N$ and $\mathcal{F}$ be the Fermionic Fock space over $\ell_2(\Lambda_N)$. If $A$ is a linear operator on $\ell_2(\Lambda_N)$, then $\Gamma(A)$ denotes its second quantization as an operator on $\mathcal{F}$. Let $P^N_-$ be the projection onto $[-N,\ldots,0]$ and let $\Omega_N$ be the corresponding Slater determinant. We set, as operators in $\ell_2$,
\begin{equation}\label{B.4}
(T^q_+)_{ij}=
 \begin{cases}
 (\sqrt{q})^{i-j} & \textrm{for }i\geq j,\\
 0 & \textrm{for }i<j,
 \end{cases}
\end{equation}
and $T^q_-=(T^q_+)^*$. $T^{q,N}_+$, $T^{q,N}_-$ is the restriction of $T^q_+$, $T^q_-$ to $\ell_2(\Lambda_N)$. Finally let $a(j)$ be the Fermion field with index $j$, $|j|\leq N$. Then
\begin{equation}\label{B.5}
K^q_{m,d}(i,j)=\lim_{N\to\infty}\frac{1}{Z_N}\langle\Omega_N,
\Gamma(T^{q,N}_- T^{q,N}_+)^{m+d}
a^*(j)a(i)\Gamma(T^{q,N}_- T^{q,N}_+)^{m-d}\Omega_N\rangle_\mathcal{F}
\end{equation}
with the normalization
\begin{equation}\label{B.6}
Z_N=\langle \Omega_N,\Gamma(T^{q,N}_- T^{q,N}_+)^{2m}
\Omega_N\rangle_\mathcal{F}.
\end{equation}
Here $\langle\cdot,\cdot\rangle_\mathcal{F}$ denotes the inner product in Fock space. Working out the limit, see~\cite{PS02}, results in
\begin{equation}\label{B.7}
K^q_{m,d}(i,j)=\big((T^q_+)^{m+d}(T^q_-)^{-(m-d)}P_-(T^q_+)^{-(m+d)}(T^q_-)^{m-d}\big)(i,j),
\end{equation}
where $P_- =\lim_{N\to\infty}P^N_-$ projects onto $\Z_-=(\ldots,-1,0]$.

In Fourier space $T^q_+$ is multiplication by
$(1-\sqrt{q}e^{-ik})^{-1}=\hat{T}^q_+(k)$ and $T^q_-$
multiplication by $(1-\sqrt{q}e^{ik})^{-1}=\hat{T}^q_-(k)$. The
rescaling in (\ref{B.3}) amounts to replacing $q$ by
$1-\e$ and $k$ by $\e k$. Then
\begin{equation}\label{B.7a}
\lim_{\e\to 0}\e \hat{T}^{1-\e}_+(\e k)=(1+\tfrac{1}{2}ik)^{-1},\quad \lim_{\e\to 0}\e \hat{T}^{1-\e}_-(\e k)=(1-\tfrac{1}{2}ik)^{-1}.
\end{equation}
By inserting the limit (\ref{B.7a}) in (\ref{B.7}), the claim
(\ref{B.3}) follows.

\subsection{Boundary sources}
We add boundary sources through the random variables $w(0,0)$, $w(j,0)$, $w(0,j)$, $j\geq 1$. They are independent and geometrically distributed according to
\begin{eqnarray}\label{B.8}
\Pb(\{w(0,0)=n\}) &=&
(1-\alpha\beta)(\alpha\beta)^n,\nonumber\\
\Pb(\{w(j,0)=n\}) &=&
(1-\alpha\sqrt{q})(\alpha\sqrt{q})^n,\nonumber\\
\Pb(\{w(0,j)=n\}) &=& (1-\beta\sqrt{q})(\beta\sqrt{q})^n,
\end{eqnarray}
with $0<\alpha\beta,\alpha\sqrt{q},\beta\sqrt{q}<1$.

The corresponding random field $\phi^q_{\tau,\alpha\beta}$ is again determinantal with defining kernel $K^{(2m+1)}_{q,\alpha\beta}$, where we set $\tau=2m+1$ in accordance with the convention of Section~\ref{sect3}. $K^{(2m+1)}_{q,\alpha\beta}$ at equal Fermionic time $2d$ is computed by the same method as in Section~\ref{sectB1}. In particular we first restrict the height lines to the interval $[-N,\ldots,N]$ and then take the limit $N\to\infty$. Let
\begin{equation}
f_\alpha(j)=\alpha^j,\quad
a(f_\alpha)=\sum_{j\in\Z}f_\alpha(j)a(j)
\end{equation}
and let $\Omega^-_N$ be the ground state vector with sites $-N,\ldots,-1$ filled and sites $0,\ldots,N$ empty. Then $a^*(f_\alpha)\Omega^-_N$ gives the correct weight to the jump at the right boundary, correspondingly for the left boundary. Therefore the defining kernel is given through
\begin{eqnarray}\label{B.9}
K^{(2m+1)}_{q,\alpha\beta}(2d,i;2d,j) &=&
\lim_{N\to\infty}\frac{1}{Z_N}\langle \Omega^-_N,a(f_\beta)
\Gamma(T^{q,N}_- T^{q,N}_+)^{m+d}\\
& & a^*(i)a(j)\Gamma(T^{q,N}_- T^{q,N}_+)^{m-d} a^*(f_\alpha)\Omega^-_N\rangle_\mathcal{F}\nonumber
\end{eqnarray}
with the normalization
\begin{equation}\label{B.10}
Z_N= \langle\Omega^-_N,a(f_\beta)\Gamma(T^{q,N}_-T^{q,N}_+)^{2m} a^*(f_\alpha)\Omega^-_N\rangle_\mathcal{F}.
\end{equation}
We note that
\begin{equation}\label{B.12}
\Gamma(A)a^*(f)= a^*(Af)\Gamma(A)
\end{equation}
and
\begin{eqnarray}\label{B.13}
T^q_+ f_\alpha &=& (1-\alpha^{-1}\sqrt{q})^{-1} f_\alpha,\quad
\sqrt{q} <\alpha,\nonumber\\
T^q_- f_\beta &=& (1-\beta\sqrt{q})^{-1}f_\beta ,\quad
\beta<\frac{1}{\sqrt{q}}.
\end{eqnarray}
Therefore
\begin{eqnarray}\label{B.14}
K^{(2m+1)}_{q,\alpha\beta}(2d,i;2d,j) &=&
\lim_{N\to\infty}\frac{1}{\tilde Z_{N,\alpha\beta}}
\langle \Omega^-_N,\Gamma(T^{q,N}_- T^{q,N}_+)^{m+d}\\
& & a(f_\beta)a^*(i)a(j)a^*(f_\alpha)\Gamma(T^{q,N}_- T^{q,N}_+)^{m-d} \Omega^-_N\rangle_\mathcal{F}\nonumber
\end{eqnarray}
with
\begin{equation}
\tilde Z_{N,\alpha\beta} = \langle \Omega^-_N,\Gamma(T^{q,N}_- T^{q,N}_+)^{m+d} a(f_\beta)a^*(f_\alpha)\Gamma(T^{q,N}_- T^{q,N}_+)^{m-d} \Omega^-_N\rangle_\mathcal{F}.
\end{equation}
Following the steps of Section~\ref{sectB1}, one concludes that
\begin{eqnarray}
& &\lim_{N\to\infty} \tilde Z_{N,\alpha\beta}= \tilde Z_{\alpha\beta} = (\alpha\beta)^{-1} \langle f_\alpha, (\Id-K^q_{m,d}) f_\beta\rangle_{\ell_2} \\
&&= (1-\alpha\beta)^{-1} \left(\frac{(1-\sqrt{q}/\beta)(1-\sqrt{q}\beta)}{(1-\sqrt{q}/\alpha)(1-\sqrt{q}\alpha)}\right)^d\left(\frac{(1-\sqrt{q}/\alpha)(1-\sqrt{q}/\beta)}{(1-\sqrt{q}\alpha)(1-\sqrt{q}\beta)}\right)^m.\nonumber
\end{eqnarray}

At this stage it is of use to recall a general property of quasifree states on CAR-algebras. Let $\cal A$ be a CAR algebra indexed by $\Z$ and let $\omega$ be a quasifree linear functional (a state) on $\cal A$, uniquely defined through
\begin{equation}
\omega(a(j))=\omega(a^*(j))=0,\quad \omega(a^*(j) a(j))=K(i,j)
\end{equation}
with $K$ a positive linear operator on $\ell_2$, $\|K\|\leq 1$. Let $f\in \ell_2$ and $a(f)=\sum_{j\in\Z} f(j)a(j)$. For $f_\ell,f_r\in\ell_2$ we define
\begin{equation}
Z=\omega(a(f_\ell)a^*(f_r))=\langle f_r, (\Id-K)f_\ell\rangle_{\ell_2} < \infty
\end{equation}
and a linear functional $\tilde\omega$ through
\begin{equation}\label{B.20}
\tilde\omega(A)=\frac{1}{Z}\omega(a(f_\ell) A a^*(f_r)),\quad A\in {\cal A}.
\end{equation}
Then $\tilde\omega(\Id)=1$ and $\tilde\omega$ is again quasifree with $\tilde\omega(a(j))=\tilde\omega(a^*(j))=0$ and covariance
\begin{equation}
\tilde\omega(a^*(i)a(j))=K(i,j)+\frac{1}{Z}(\Id-K)f_\ell(i) (\Id-K)^* f_r(j).
\end{equation}

With the results from Section~\ref{sectB1}, the limit state (\ref{B.14}) is a quasifree linear functional precisely of the form (\ref{B.20}) with
\begin{equation}
K(i,j)=K^q_{m,d}(i+1,j+1),
\end{equation}
see (\ref{B.7}), and $f_\ell(j)=\beta^j$, $f_r(j)=\alpha^j$, $\sqrt{q} < \alpha,\beta < \sqrt{q}^{-1}$.

We take the exponential limit by choosing
\begin{equation}
q=1-\e,\quad \alpha=1-\e a,\quad \beta=1-\e b,\quad -\frac12 < a,b < \frac12, \quad a+b>0.
\end{equation}
Then
\begin{equation}
\lim_{\e\to 0} f_{1-\e a}(\lfloor x/\e\rfloor ) = \psi_a(x)=e^{-a x}
\end{equation}
and, for $x,y>0$,
\begin{eqnarray}
& & \lim_{\e\to 0} \e^{-1} K^{(2m+1)}_{1-\e,\alpha\beta}\big(2d,\lfloor \e^{-1}x\rfloor; 2d,\lfloor \e^{-1} y \rfloor\big)\\
&&= K_{m,d}(x,y) + \frac{1}{Z_{a,b}}(\Id-K_{m,d})\psi_b(x) (\Id-K_{m,d})^*\psi_a(y)\nonumber
\end{eqnarray}
with $Z_{a,b}$ given in (\ref{4.14}). This completes the proof of Proposition~\ref{Prop4}.

\subsection{Proof of Proposition~\ref{PropInverse}}
\begin{proofOF}{Proposition~\ref{PropInverse}}
By Proposition~\ref{PropLaguerre} $K_{m,d}$ is a similarity transform of the Laguerre kernel. Therefore $K_{m,d}=(K_{m,d})^2$, $\|K_{m,d}\|=1$, $P_u K_{m,d} P_u$ is trace class, and all eigenvalues of $P_u K_{m,d} P_u$ are in the interval $[0,1]$. Thus we only prove that, if $u>0$, $1$ is not in the spectrum of $P_u K_{m,d} P_u$, which is accomplished by \emph{reductio ad absurdum}. Assume that $\psi\in L^2(\R_+)$ is an eigenfunction for eigenvalue $1$,
\begin{equation}
P_u K_{m,d} P_u \psi = \psi.
\end{equation}
Then $\psi(x)=0$ for $x\in [0,u)$. On the other hand,
\begin{equation}
\|\psi\| \leq \| K_{m,d} P_u \psi\| \leq \| \psi \|.
\end{equation}
Hence $\|K_{m,d} P_u \psi\| = \| \psi \|$ and, since $\|K_{m,d} P_u\|=1$, one concludes that
\begin{equation}
K_{m,d} P_u \psi = \psi.
\end{equation}
Therefore $\psi$ is of the form $(\textrm{finite polynomial})\times e^{-x/2}$, which cannot vanish identically on $[0,u)$. Thus the contradiction.

To establish the second claim we use that $\psi_a$ is eigenfunction of $R$ to obtain
\begin{equation}
(P_u(\Id-K_{m,d})\psi_a)(x) = \Theta(x-u)\Big(e^{-ax}-Z(-a) \int_{\R_-}\dx y L(x,y) \psi_a(y) \Big).
\end{equation}
Moreover, it is easy to see that $|I_{m,d}(z)|\leq 2\pi C_{m,d} e^{-\beta z}$ for any $0<\beta<1-\rho$, $C_{m,d}$ being a constant (take as path $\Gamma_{1-\rho}$ the circle centered in $1-\rho$ of radius $1-\rho-\beta$). Let us choose any $\beta\in (1/2-\rho+a,1-\rho)$. Then, for $u>0$,
\begin{eqnarray}
& &\big|(P_u(\Id-K_{m,d})\psi_a)(x)\big| \leq \Theta(x) e^{-ax}+\\
& &+\Theta(x) Z(-a) C_{m,d} e^{-(\beta-\frac12+\rho)x} \int_{\R_-}\dx y e^{y(\beta+\rho-\frac12-a)}. \nonumber
\end{eqnarray}
$\beta-\frac12+\rho>0$, because $a>0$, and $\beta+\rho-\frac12-a>0$, because $\beta>\frac12-\rho+a$. Thus $P_u(\Id-K_{m,d})\psi_a\in L^2(\R)$ with norm uniformly bounded in $u$. The second part of (\ref{eq3.22b}) is treated similarly.
$\psi_a$ is eigenfunction of $L^*$, thus
\begin{equation}
(P_u(\Id-K_{m,d})^*\psi_a)(x) = \Theta(x-u)\Big(e^{-ax}-Z(a)^{-1} \int_{\R_-}\dx y R(y,x) e^{-a y} \Big).
\end{equation}
$|\tilde I_{m,d}(z)|\leq 2\pi \tilde C_{m,d} e^{-\beta z}$ for any $0<\beta<\rho$. Thus, by choosing $\beta\in (\rho+a-1/2,\rho)$, for $u>0$,
\begin{eqnarray}
& &\big|(P_u(\Id-K_{m,d})\psi_a)(x)\big| \leq \Theta(x) e^{-ax}+\\
& &+\Theta(x) Z(-a) \tilde C_{m,d} e^{-(\beta+\frac12-\rho)x} \int_{\R_-}\dx y e^{y(\beta-\rho+\frac12-a)}, \nonumber
\end{eqnarray}
which, by the choice of $\beta$ and since $a>0$, implies $P_u(\Id-K_{m,d})^*\psi_a\in L^2(\R)$ with norm uniformly bounded in $u$.
\end{proofOF}

\section{The Laguerre kernel}\label{AppLaguerre}
Let $L_n^{(\alpha)}$ be the standard $n$-th Laguerre polynomial of integer order $\alpha$, $\alpha\geq 0$ \cite{MOS66}, Chapter 5.3. The Laguerre polynomials are orthogonal on $\R_+$ relative to the weight $x^\alpha e^{-x}$ as
\begin{equation}\label{C.1}
\int_{\R_+}\dx x x^{\alpha} e^{-x} \left(\frac{n!}{(n+\alpha)!}\right)^{1/2} L_n^{(\alpha)}(x) \left(\frac{m!}{(m+\alpha)!}\right)^{1/2} L_m^{(\alpha)}(x) =\delta_{n,m}.
\end{equation}
The Laguerre kernel is the orthogonal projection onto the first $n$ Laguerre polynomials and is given by
\begin{equation}\label{C.2}
K_n^{(\alpha)}(x,y) = \sum_{j=0}^{n-1} \frac{j!}{(j+\alpha)!} L_j^{(\alpha)}(x) L_j^{(\alpha)}(y) x^{\alpha/2} y^{\alpha/2} e^{-x/2} e^{-y/2}
\end{equation}
for $x,y\geq 0$.

\begin{proposition}\label{PropLaguerre}
With definitions (\ref{4.11c}) and (\ref{C.2}) one has
\begin{equation}\label{CP1}
K_{m,-d}(x,y)=K_{m,d}(x,y)
\end{equation}
and
\begin{equation}\label{CP2}
K_{m,d}(x,y)=K_{m-d}^{(2d)}(x,y) \Big(\frac{x}{y}\Big)^d
\end{equation}
for $0\leq d < m$ and $x,y >0$.
\end{proposition}
\begin{proof}
Let us define
\begin{eqnarray}
\hat g_{\ell}(k)&=&\big(\tfrac12-ik\big)^{m-d}\big(\tfrac12+ik\big)^{-(m+d)},\nonumber \\
\hat g_{r}(k)&=&\big(\tfrac12-ik\big)^{m+d}\big(\tfrac12+ik\big)^{-(m-d)}.
\end{eqnarray}
We set $\alpha=2d$, $n=m-d-1$. By \cite{MOS66}, p.\ 244, it follows
\begin{equation}\label{C.3}
\int_{\R}\dx k e^{i k x}\hat g_{\ell}(k)=g_{\ell}(x) = (-1)^n \frac{n!}{(n+\alpha)!} e^{-x/2}(1-\partial_x) x^\alpha L_n^{(\alpha)}(x)\Theta(x)
\end{equation}
and
\begin{equation}
\int_{\R}\dx k e^{-i k x}\hat g_{r}(k)=g_{r}(-x),
\end{equation}
with
\begin{equation}\label{C.4}
g_r(x)= \big(\tfrac12+\partial_x\big)^d \big(\tfrac12 -\partial_x\big)^{d+1}(-1)^{m+1} e^{-x/2} L_{m-1}^{(0)}(x)\Theta(x).
\end{equation}
$g_\ell$ is supported on $[0,\infty)$ with a discontinuity at $x=0$. As a distribution $g_r$ is supported in $[0,\infty)$ with the singular part concentrated at $\{x=0\}$.

By \cite{MOS66}, Section 5.5.2, one has the identities
\begin{equation}
(1-\partial_x)L_n^{(\alpha)} = -\partial_x L_{n+1}^{(\alpha)},\quad \partial_x L_n^{(\alpha)} = - L_{n+1}^{(\alpha+1)}.
\end{equation}
Using them repeatedly in (\ref{C.3}) and (\ref{C.4}) yields, for $x>0$,
\begin{equation}\label{C.6}
g_\ell(x)=(-1)^{n+1} \frac{n!}{(n+\alpha)!} e^{-x/2}\big(x^\alpha \partial_x L_{n+1}^{(\alpha)}(x)+\alpha x^{\alpha-1} L_n^{(\alpha)}\big),
\end{equation}
and
\begin{equation}\label{C.7}
g_r(x)=(-1)^{n+1} e^{-x/2} \partial_x L_{n+1}^{(\alpha)}(x).
\end{equation}

With these notations the integral kernel from (\ref{4.11c}) is expressed as
\begin{eqnarray}\label{C.8}
& &\big(T_+^{m+d} T_-^{-(m-d)} P_- T_+^{-(m+d)} T_-^{m-d}\big)(x,y)  \\
&&= \int_{\R_-}\dx w g_\ell(x-w)g_r(-(w-y))=\int_{\R_+}\dx w g_\ell(x+w)g_r(y+w).\nonumber
\end{eqnarray}
Note that for $x,y>0$ only the regular part of $g_r$ is used. We insert (\ref{C.6}) and (\ref{C.7}) in (\ref{C.8}). Then (\ref{CP2}) of Proposition~\ref{PropLaguerre} amounts to
\begin{eqnarray}\label{C.9}
 & &\sum_{j=0}^n \frac{j!}{(j+\alpha)!} x^\alpha L_j^{(\alpha)}(x) L_j^{(\alpha)}(y) =
\frac{n!}{(n+\alpha)!} \int_{\R_+}\dx w e^{-w} \big(\partial_w L_{n+1}^{(\alpha)}(x+w)\big) \nonumber \\
& & \times \big((x+w)^\alpha \partial_w L_{n+1}^{(\alpha)}(x+w)+\alpha (x+w)^{\alpha-1} L_n^{(\alpha)}(y+w)\big).
\end{eqnarray}

We check recursively by setting the left hand side as $\sum_{j=0}^n j!/(j+\alpha)! B_j$ and the right side as $n!/(n+\alpha)! A_n$. Then (\ref{C.9}) is equivalent to
\begin{equation}\label{C.10}
A_0=B_0,\quad \frac{n!}{(n+\alpha)!} A_n-\frac{(n-1)!}{(n-1+\alpha)!} A_{n-1}=\frac{n!}{(n+\alpha)!}B_n,\quad n=1,2,\ldots.
\end{equation}
$A_0=B_0$ amounts to a partial integration. For the second equality we write
\begin{eqnarray}
& &A_n-\frac{n+\alpha}{n}A_{n-1} = \int_{\R_+}\dx w e^{-w} (x+w)^\alpha \Big(L_{n+1}'(x+w)L_{n+1}'(y+w) \nonumber \\
& &-\frac{n+\alpha}{n}L_{n}'(x+w)L_{n}'(y+w)\Big) + \int_{\R_+}\dx w e^{-w} \alpha (x+w)^{\alpha-1} \nonumber \\
& &\times\Big(L_{n}(x+w)L_{n+1}'(y+w)-\frac{n+\alpha}{n}L_{n-1}(x+w)L_{n}'(y+w)\Big),
\end{eqnarray}
omitting the superscript $\alpha$. In the second integral we use the identities $L_{n+1}'=L_n'-L_n$ and $x L_n'=n L_n-(n+\alpha)L_{n-1}$. Then the terms combine as
\begin{eqnarray}
A_n-\frac{n+\alpha}{n}A_{n-1} &=& \int_{\R_+}\dx w e^{-w} (x+w)^\alpha \big(L_n(x+w)L_n(y+w) \nonumber \\
& & - L_n'(x+w)L_n(y+w)-L_n(x+w)L_n'(y+w)\big) \nonumber \\
& & - \int_{\R_+}\dx w e^{-w} \alpha (x+w)^{\alpha-1} L_n(x+w) L_n(y+w) \nonumber \\
&=& -\int_{\R_+}\dx w \Dt{}{w}\big(e^{-w}(x+w)^\alpha L_n(x+w) L_n(y+w)\big)\nonumber \\
&=& B_n,
\end{eqnarray}
which is the recursion relation (\ref{C.10}).
\end{proof}

\section{Some useful relations}
\subsection{Two representations of the Airy functions}
For any $\sigma<0$, define the path $\gamma_\sigma=\sigma+ i \R$. Then
\begin{equation}\label{Airy1}
\Ai(z)=\frac{1}{2\pi i} \int_{\gamma_\sigma} \dx \xi e^{-\frac13 \xi^3 +z \xi}
\end{equation}
and for any $\mu>0$, with $\gamma_\mu=\mu+ i \R$,
\begin{equation}\label{Airy2}
\Ai(z)=\frac{1}{2\pi i} \int_{\gamma_\mu} \dx \xi e^{\frac13 \xi^3 -z \xi}.
\end{equation}
One can deform $\gamma_\sigma$ so that it goes from $\infty e^{-2\pi i/3}$ to $\infty e^{2\pi i/3}$ and crosses the real axis at $\sigma$. In this case the paths will be denoted by $\gamma_\sigma^{\ell}$. Similarly the deformation of $\gamma_\mu$ goes from $\infty e^{-i\pi/3}$ to $\infty e^{i\pi/3}$ and is denoted by $\gamma_\mu^r$.

A formula which will be employed later is
\begin{equation}\label{Airy3}
M(w)=\int_\R \dx y e^{w y} \Ai(\beta+y) = e^{\frac13 w^3-\beta w}
\end{equation}
valid for all $w\geq 0$ (for $w=0$ as improper Riemann integral). To prove it one derives the differential equation
\begin{equation}\label{eqAiryProva}
\Dt{M(w)}{w} = M(w)(w^2-\beta)
\end{equation}
by integrating twice by parts. (\ref{Airy3}) follows from (\ref{eqAiryProva}) and the initial value $M(w=0)=1$.

\subsection{Two integrals around a pole}
Let $\Gamma_w$ be the path enclosing $z=w$ and anti-clockwise oriented. Then
\begin{equation}\label{eq122}
\int_{\Gamma_w}\dx \xi \frac{e^{\frac13 \xi^3-\frac13 w^3 + (w^2+s)(w-\xi)}}{2\pi i (w-\xi)^2} = \textrm{Res}\bigg(\frac{e^{\frac13 \xi^3-\frac13 w^3 + (w^2+s)(w-\xi)}}{(w-\xi)^2};\xi=w\bigg) = -s
\end{equation}
and
\begin{equation}
\int_{\Gamma_w}\dx \xi \frac{e^{-\frac13 \xi^3+\frac13 w^3 + (w^2+s)(\xi-w)}}{2\pi i (\xi-w)^2} = \textrm{Res}\bigg(\frac{e^{-\frac13 \xi^3+\frac13 w^3 + (w^2+s)(\xi-w)}}{(\xi-w)^2};\xi=w\bigg) = s.
\end{equation}

\subsection{Two equivalent expressions for $S_{w,s}$}\label{Sws}
Let us consider $w>0$. The representation of the Airy function (\ref{Airy2}) allows one to write
\begin{eqnarray}
S_{w,s}&=&s+e^{-\frac13 w^3} \int_0^\infty \dx x \int_0^\infty \dx y \Ai(w^2+s+x+y) e^{w(w^2+s+x+y)} \nonumber\\
&=& s+e^{-\frac13 w^3} \int_0^\infty \dx x \int_0^\infty \dx y \frac{1}{2\pi i} \int_{\gamma_\mu^r} \dx \xi e^{\frac13 \xi^3} e^{(w^2+s+x+y)(w-\xi)}
\end{eqnarray}
By choosing $\mu=2w$, one obtains an integrand which is absolutely integrable. Thus by Fubini's Theorem we are allowed to exchange the order of the integrals and compute first the one in $(x,y)$ with the result
\begin{equation}
S_{w,s}= s+\int_{\gamma_{2w}^r} \dx \xi \frac{e^{\frac13 \xi^3-\frac13 w^3 + (w^2+s)(w-\xi)}}{2\pi i (w-\xi)^2}.
\end{equation}
Let $\Gamma_w$ be as in (\ref{eq122}), then
\begin{eqnarray}
S_{w,s}&=&s+\int_{\gamma_{w/2}^r} \dx \xi \frac{e^{\frac13 \xi^3-\frac13 w^3 + (w^2+s)(w-\xi)}}{2\pi i (w-\xi)^2} + \int_{\Gamma_w} \dx \xi \frac{e^{\frac13 \xi^3-\frac13 w^3 + (w^2+s)(w-\xi)}}{2\pi i (w-\xi)^2} \nonumber \\
&=& \int_{\gamma_{w/2}^r} \dx \xi \frac{e^{\frac13 \xi^3-\frac13 w^3 + (w^2+s)(w-\xi)}}{2\pi i (w-\xi)^2}.
\end{eqnarray}
On the other hand,
\begin{eqnarray}
\int_{\R_-^2}\dx x \dx y \Phi_{w,s}(x+y) &=& \int_{\R_-^2}\dx x \dx y \frac{1}{2\pi i} \int_{w/2+i\R}\dx \xi e^{\frac13 \xi^3-\frac13 w^3}e^{(w-\xi)(w^2+s+x+y)} \nonumber \\
&=& \int_{w/2+i\R}\dx \xi \frac{e^{\frac13 \xi^3-\frac13 w^3 + (w^2+s)(w-\xi)}}{2\pi i (w-\xi)^2} = S_{w,s}
\end{eqnarray}
by deforming the integral on $w/2+i\R$ to $\gamma_{w/2}^r$.


\begin{thebibliography}{10}
\bibitem{BBP04}
J.~Baik, G.~Ben Arous, and S.~P\'ech\'e, \emph{Phase transition of the largest
  eigenvalue for non-null complex sample covariance matrices}, Ann. Probab.
  \textbf{33} (2005), 1643--1697.

\bibitem{BR00}
J.~Baik and E.M. Rains, \emph{Limiting distributions for a polynuclear growth
  model with external sources}, J. Stat. Phys. \textbf{100} (2000), 523--542.

\bibitem{BR99}
J.~Baik and E.M. Rains, \emph{Symmetrized random permutations}, Random Matrix
  Models and Their Applications, vol.~40, Cambridge University Press, 2001,
  pp.~1--19.

\bibitem{BKS85}
H.~van Beijeren, R.~Kutner, and H.~Spohn, \emph{Excess noise for driven
  diffusive systems}, Phys. Rev. Lett. \textbf{54} (1985), 2026--2029.

\bibitem{CM02}
F.~Colaiori and M.A. Moore, \emph{Numerical solution of the mode-coupling
  equations for the {Kardar-Parisi-Zhang} equation in one dimension}, Phys.
  Rev. E \textbf{65} (2002), 017105.

\bibitem{Fer04}
P.L. Ferrari, \emph{Polynuclear growth on a flat substrate and edge scaling of
  {GOE} eigenvalues}, Comm. Math. Phys. \textbf{252} (2004), 77--109.

\bibitem{FS05b}
P.L. Ferrari and H.~Spohn, \emph{{A determinantal formula for the GOE
  Tracy-Widom distribution}}, J. Phys. A \textbf{38} (2005), L557--L561.

\bibitem{FNS77}
D.~Forster, D.R. Nelson, and M.J. Stephen, \emph{Large-distance and long-time
  properties of a randomly stirred fluid}, Phys. Rev. A \textbf{16} (1977),
  732--749.

\bibitem{SI04}
T.~Imamura and T.~Sasamoto, \emph{Fluctuations of the one-dimensional
  polynuclear growth model with external sources}, Nucl. Phys. B \textbf{699}
  (2004), 503--544.

\bibitem{Jo00b}
K.~Johansson, \emph{Shape fluctuations and random matrices}, Comm. Math. Phys.
  \textbf{209} (2000), 437--476.

\bibitem{Jo03b}
K.~Johansson, \emph{Discrete polynuclear growth and determinantal processes},
  Comm. Math. Phys. \textbf{242} (2003), 277--329.

\bibitem{KPZ86}
K.~Kardar, G.~Parisi, and Y.Z. Zhang, \emph{Dynamic scaling of growing
  interfaces}, Phys. Rev. Lett. \textbf{56} (1986), 889--892.

\bibitem{KMH92}
J.~Krug, P.~Meakin, and T.~Halpin-Healy, \emph{Amplitude universality for
  driven interfaces and directed polymers in random media}, Phys. Rev. A
  \textbf{45} (1992), 638--653.

\bibitem{Lig76}
T.M. Liggett, \emph{Coupling the simple exclusion process}, Ann. Probab.
  \textbf{4} (1976), 339--356.

\bibitem{Li99}
T.M. Liggett, \emph{Stochastic interacting systems: contact, voter and
  exclusion processes}, Springer Verlag, Berlin, 1985.

\bibitem{MOS66}
W.~Magnus, F.~Oberhettinger, and R.P. Soni, \emph{Formulas and theorems for the
  special functions of mathematical physics}, Grundlehren Band 52, Springer
  Verlag, Berlin, 1966.

\bibitem{Ok01}
A.~Okounkov, \emph{Infinite wedge and random partitions}, Selecta Math.
  \textbf{7} (2001), 57--81.

\bibitem{Pra03}
M.~Pr{\"a}hofer, \emph{Stochastic surface growth}, Ph.D. thesis,
  Ludwig-Maximilians-Universit{\"a}t, M{\"u}nchen,
  http://edoc.ub.uni-muenchen.de/archive/00001381, 2003.

\bibitem{PS01}
M.~Pr{\"a}hofer and H.~Spohn, \emph{Current fluctuations for the totally
  asymmetric simple exclusion process}, In and out of equilibrium
  (V.~Sidoravicius, ed.), Progress in Probability, Birkh{\"a}user, 2002.

\bibitem{PS02}
M.~Pr{\"a}hofer and H.~Spohn, \emph{Scale invariance of the {PNG} droplet and
  the {A}iry process}, J. Stat. Phys. \textbf{108} (2002), 1071--1106.

\bibitem{PS02b}
M.~Pr{\"a}hofer and H.~Spohn, \emph{Exact scaling function for one-dimensional
  stationary {KPZ} growth}, J. Stat. Phys. \textbf{115} (2004), 255--279.

\bibitem{Sas05}
T.~Sasamoto, \emph{Spatial correlations of the {1D KPZ} surface on a flat
  substrate}, J. Phys. A \textbf{38} (2005), L549--L556.

\bibitem{Sp83}
H.~Spohn, \emph{Excess noise for a lattice gas model of a resistor}, Z. Phys. B
  \textbf{57} (1984), 255--261.

\bibitem{TW94}
C.A. Tracy and H.~Widom, \emph{Level-spacing distributions and the {A}iry
  kernel}, Comm. Math. Phys. \textbf{159} (1994), 151--174.

\end{thebibliography}
\end{document}